\documentclass{article}
\usepackage[utf8]{inputenc}

\usepackage{graphicx}
\usepackage[english]{babel}
\usepackage{authblk} 

\usepackage{comment}

\usepackage[numbers]{natbib}
\usepackage{hyperref}       
\usepackage{url}            
\usepackage{booktabs}       
\usepackage{amsfonts}       
\usepackage{nicefrac}       
\usepackage{microtype}      
\usepackage{wrapfig}
\usepackage{graphicx}
\usepackage[inline]{enumitem}
\usepackage{bm}
\usepackage{amsfonts,amssymb,amsmath}
\usepackage[dvipsnames]{xcolor}
\usepackage{graphicx}
\usepackage{subcaption}
\usepackage[normalem]{ulem}
\usepackage{lineno}

\newcommand{\beginsupplement}{%
        \setcounter{table}{0}
        \renewcommand{\thetable}{S\arabic{table}}%
        \setcounter{figure}{0}
        \renewcommand{\thefigure}{S\arabic{figure}}%
        \renewcommand\thesection{S}%
        \renewcommand\thesubsection{S.\arabic{subsection}}%
     }
     
\title{Overcoming Barriers to Data Sharing with Medical Image Generation: A Comprehensive Evaluation}

\author[1,5,*]{August DuMont Schütte}
\author[2,3]{Jürgen Hetzel}
\author[4]{Sergios Gatidis}
\author[4,5]{Tobias Hepp}
\author[1]{Benedikt Dietz}
\author[5,6]{Stefan Bauer}
\author[7]{Patrick Schwab}
\affil[1]{ETH Zurich, Switzerland}
\affil[2]{Department of Medical Oncology and Pneumology, University Hospital of Tübingen, Germany}
\affil[3]{Department of Pneumology, Kantonsspital Winterthur, Switzerland}
\affil[4]{Department of Radiology, University Hospital of Tübingen, Germany}
\affil[5]{Max Planck Institute for Intelligent Systems, Tübingen, Germany}
\affil[6]{CIFAR Azrieli Global Scholar}
\affil[7]{GlaxoSmithKline, Artificial Intelligence \& Machine Learning, Switzerland}
\affil[*]{Corresponding author, \href{mail}{augustschdmnt@gmail.com}}
\date{\today}                  
\begin{document}

\maketitle

\section{Abstract}

Privacy concerns around sharing personally identifiable information are a major practical barrier to data sharing in medical research. However, in many cases, researchers have no interest in a particular individual's information but rather aim to derive insights at the level of cohorts. Here, we utilize Generative Adversarial Networks (GANs) to create derived medical imaging datasets consisting entirely of synthetic patient data. The synthetic images ideally have, in aggregate, similar statistical properties to those of a source dataset but do not contain sensitive personal information. We assess the quality of synthetic data generated by two GAN models for chest radiographs with $14$ different radiology findings and brain computed tomography (CT) scans with six types of intracranial hemorrhages. We measure the synthetic image quality by the performance difference of predictive models trained on either the synthetic or the real dataset. We find that synthetic data performance disproportionately benefits from a reduced number of unique label combinations. Our open-source benchmark also indicates that at low number of samples per class, label overfitting effects start to dominate GAN training. We show that synthetic data generation can benefit from increasing the spatial resolution up to $128 \times 128$ pixels, after which training instabilities make the generation of realistic imaging data more difficult. We additionally conducted a reader study in which trained radiologists do not perform better than random on discriminating between synthetic and real medical images for both data modalities to a statistically significant extent for intermediate levels of resolutions. In accordance with our benchmark results, radiologists can discriminate more accurately between real and synthetic chest radiographs at higher spatial resolution due to the emergence of visual artifacts and fine-scaled details. Our study offers valuable guidelines and outlines practical conditions under which insights derived from synthetic medical images are similar to those that would have been derived from real imaging data. Our results indicate that synthetic data sharing may be an attractive and privacy-preserving alternative to sharing real patient-level data in the right settings.

\section{Introduction}
Sharing sensitive data under strict privacy regulations remains a crucial challenge in advancing medical research \cite{lo2015}. By accessing large amounts of collected data, there have been impressive research results in a range of medical fields such as genetics \cite{sanna19}, radiomics \cite{lihui16, sun2018}, neuroscience \cite{miller16},  diagnosis \cite{DeFauw2018, Monteiro2020, Liu2020}, patient outcome prediction \cite{Courtiol2019, Matsuo2019} or drug discovery \cite{Chen2018, Bogdan19}. Particularly deep learning systems, composed of millions of trainable parameters, require large amounts of data to learn meaningful representations robustly \cite{Lecun2015}. Aside from quantity, the quality of the available patient-level data is particularly essential for medical research \cite{Esteva2019}. Highly diverse and well-curated training data empowers researchers to produce generalisable insights and reduces the risk of biased predictions when applied in practice. 

It is especially difficult to share and distribute medical data due to privacy concerns and the potential abuse of personal information \cite{Haas2011}. To overcome these privacy concerns there has been an impressive number of large-scale research collaborations to pool and curate de-identified medical data for open-source research purposes \cite{Bycroft2018, Clark2013, Tomczak2015}. Nevertheless, most medical data is still isolated and locally stored in hospitals and laboratories due to the concerns associated with sharing patient data \cite{Panhuis2010}. In many countries, privacy laws inhibit medical data sharing \cite{mark2018}, and potentially available de-identification methods lack guarantees as de-identified data can, in some cases, be linked back to individuals \cite{Na2018, Nwankwo2012}. 

In medical research, information is often analysed at the level of cohorts rather than individuals. A potential solution to the medical data sharing bottleneck, is therefore, the generation of synthetic patient data that, in aggregate, has similar statistical properties to those of a source dataset without revealing sensitive private information about individuals. While synthetic data can be generated for all kinds of data modalities, we focus on the particularly important medical imaging domain in this work. 

\begin{figure}[h]
\center
\includegraphics[width=1.\linewidth]{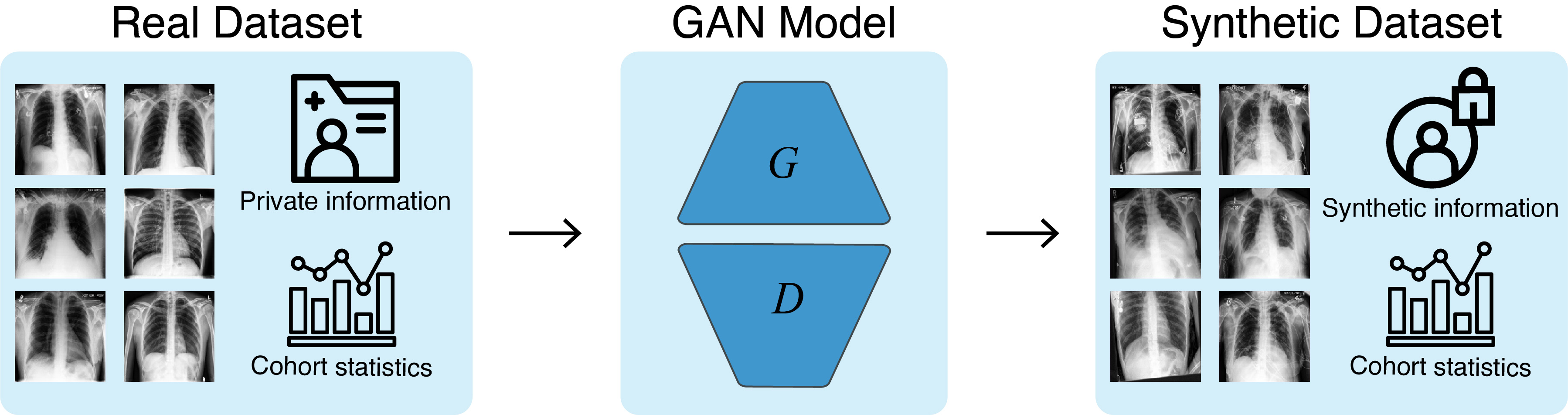}
\caption{\textbf{Synthetic medical imaging dataset generation to overcome data sharing barriers.} We train our GAN models with real medical imaging data, to generate the corresponding synthetic images. The synthetic dataset ideally no longer contains private information about individual patients while in aggregate, maintaining the real training cohort's statistical properties.}
\label{fig:1}
\end{figure}

Recently, new generative machine-learning approaches, such as Generative Adversarial Networks (GANs), have demonstrated the capability to generate realistic, high-resolution image datasets \cite{karras2018progressive}. In GANs, two neural networks play an adversarial game against each other. The generator $(G)$ tries to learn the real data distribution while the discriminator $(D)$ estimates the probability of a sample belonging to the real training set, as opposed to having been generated by $G$ \cite{goodfellow2014}. If training is stable, the model converges to a point where $D$ can no longer discriminate between real and synthetic data \cite{mescheder2017}. When each neural network is composed of a convolutional neural network (CNN), GANs have demonstrated state of the art image generation capabilities \cite{Karras2019AnalyzingAI, brock2018large}. 

Within the medical imaging domain, there are several works demonstrating the generation of realistic synthetic data, among others, retinal images \cite{costa2018, ZHAO2018}, skin lesions \cite{izadi2018generative, alceu2018, ali2019data}, hematoxylin and eosin (H\&E) stained breast cancer tissue in digital pathology \cite{quiros2020pathology}, x-ray mammographs \cite{zhou2020}, chest radiographs \cite{Chuquicusma2018, han2019breaking} and brain tumor magnetic resonance imaging (MRI) \cite{Han2020}. In \cite{med_imag_nvid} the authors illustrate the benefits of synthetic images as an additional form of data augmentation for tumor segmentation. They also analyse synthetic data sharing capabilities, but find that without fine-tuning a segmentation model on real data after it was trained on synthetic data, the performance gap is significant. The above-mentioned works develop and demonstrate GAN capabilities in specific domains, with dataset-dependent adaptations, without providing a comprehensive evaluation of how changes within and across data modalities impact different GAN model performances for synthetic data sharing. Other works such as \cite{trullo_gan} and \cite{ARMANIOUS2020101684} are less related to data sharing restrictions and instead focus on utilizing GANs for image-to-image translations within the medical domain. This idea has also been extended to semi-supervised settings where lower complexity images are synthesized first, before translating towards the higher complexity space \cite{bimodal_gan}.

Inspired by these domain-specific advancements we aim to establish a benchmark on synthetic medical imaging data generation capabilities. To the best of our knowledge, there is currently no work focused on providing a comprehensive benchmark analysis for the generation of synthetic medical images across different GAN architectures and data modalities. We offer guidelines for the use of GAN models to fully synthesise realistic datasets as a potentially viable approach to privacy-preserving data sharing, and make the following contributions:

\begin{itemize}
    \item We develop an open benchmark to analyse the generation of synthetic medical images when varying the number of label combinations, the number of samples per label combination, and the spatial resolution level present in the dataset.
    \item We present valuable guidelines for the effective generation of medical image datasets by evaluating our open-source benchmark on a reference GAN model and our newly proposed GAN architecture for two different data modalities.\footnote{The computational cost of our medical imaging benchmark amounts to approximately $32,100$ GPU-hours ($1,338$ GPU-days) on NVIDIA's Pascal P100 GPU.} 
    \item We additionally analyse privacy considerations, assess the feature importance of predictive models trained on the synthetic datasets, analyse visual artifacts at higher resolutions and finally conduct a large-scale reader study in which trained radiologists discriminate between real and synthetic medical images. 
\end{itemize}
 
\section{Results}

\subsection{Overview of approach}

Both datasets consist of binary multi-label classes. Each chest x-ray image can have a combination of the following $13$ labels: Enlarged cardiomediastinum, cardiomegaly, lung opacity,  lung lesion, edema, consolidation, pneumonia, atelectasis, pneumothorax, pleural effusion, pleural other, fracture, support device or the no finding class. The brain CT scans can consist of a combination of five different hemorrhage types: Epidural, subarachnoid, subdural, intraparenchymal and intraventricular or the no finding class. 

We randomly split each patient cohort into training, validation, and test set within strata of radiology findings, before filtering the available data for each benchmark setting. We developed all GAN models on the training datasets and stopped GAN training when the quality between real and synthetic images converged, as measured with the Fréchet Inception Distance (FID) score \cite{heusel2017gans}. Next, we generated the synthetic datasets for the train, validation, and test folds by conditioning on the labels present in the respective data folds. This means that after GAN training and inference we have a real and synthetic dataset for each benchmark setting with equivalent sizes and label combinations in all folds. In theory, a trained GAN can be used to generate unlimited amounts of data, but we want the real and synthetic folds to be equivalent for a fair comparison.

Each classifier is trained on either the real or synthetic training data fold, meaning that synthetic images are pre-computed and not generated on a batch-wise basis. In all settings, we used a pre-trained densenet-121 CNN as a predictive model, with the mean area under the receiver operating characteristics curve over all labels ($\overline{AUC}$) as the evaluation metric. For each classifier, we stopped training when the validation $\overline{AUC}$ converged. After the real predictive model is trained on the real dataset and the synthetic predictive models is trained on the synthetic dataset, we evaluated both on the separate, real data test fold to compute the difference in performance: $\overline{AUC}_{\rm real}-\overline{AUC}_{\rm syn}$. 

We repeated all experiments multiple times with varying random initialisation of the deep learning systems, allowing us to perform statistical tests on whether the distribution of $\overline{AUC}_{\rm real}-\overline{AUC}_{\rm syn}$ scores differs at different benchmark settings. Additionally, we compared the predictive models' feature importance when trained on either real or synthetic datasets. We addressed privacy concerns by analysing differences between synthetic images and the most closely matching nearest neighbour images from the entire training dataset.  Finally, we performed a large-scale reader study in which we asked trained radiologists to label a mixture of real and generated images.

\subsection{Model performance}

To accurately assess the potential of synthetic data, we analysed two model architectures across two different datasets for our benchmark. The prog-GAN model refers to the progressive GAN as a reference model, as it is still commonly used for medical image generation \cite{ali2019data, han2019breaking}. The cpD-GAN refers to our novel and improved model that we specifically developed for this benchmark. To assess the generalisation capabilities, we did not fine-tune across different benchmark settings, only when increasing the resolution, we make the necessary changes to the network architectures. 

Up to a spatial resolution of $128 \times 128$ pixels, the prog-GAN achieved an average  $\overline{AUC}_{\rm real}-\overline{AUC}_{\rm syn}$ score of $0.0495$ $(\pm 0.0276)$ across all settings on the chest radiograph dataset and $0.1367$ $(\pm 0.0324)$ across all brain scans experiments. These scores were substantially improved with the cpD-GAN that achieves $0.0206 \ (\pm 0.0100)$ on the chest x-ray settings and $0.0650 \ (\pm 0.0198)$ on the brain hemorrhage dataset experiments.

\begin{figure}[p]
\center
\includegraphics[width=1.\linewidth]{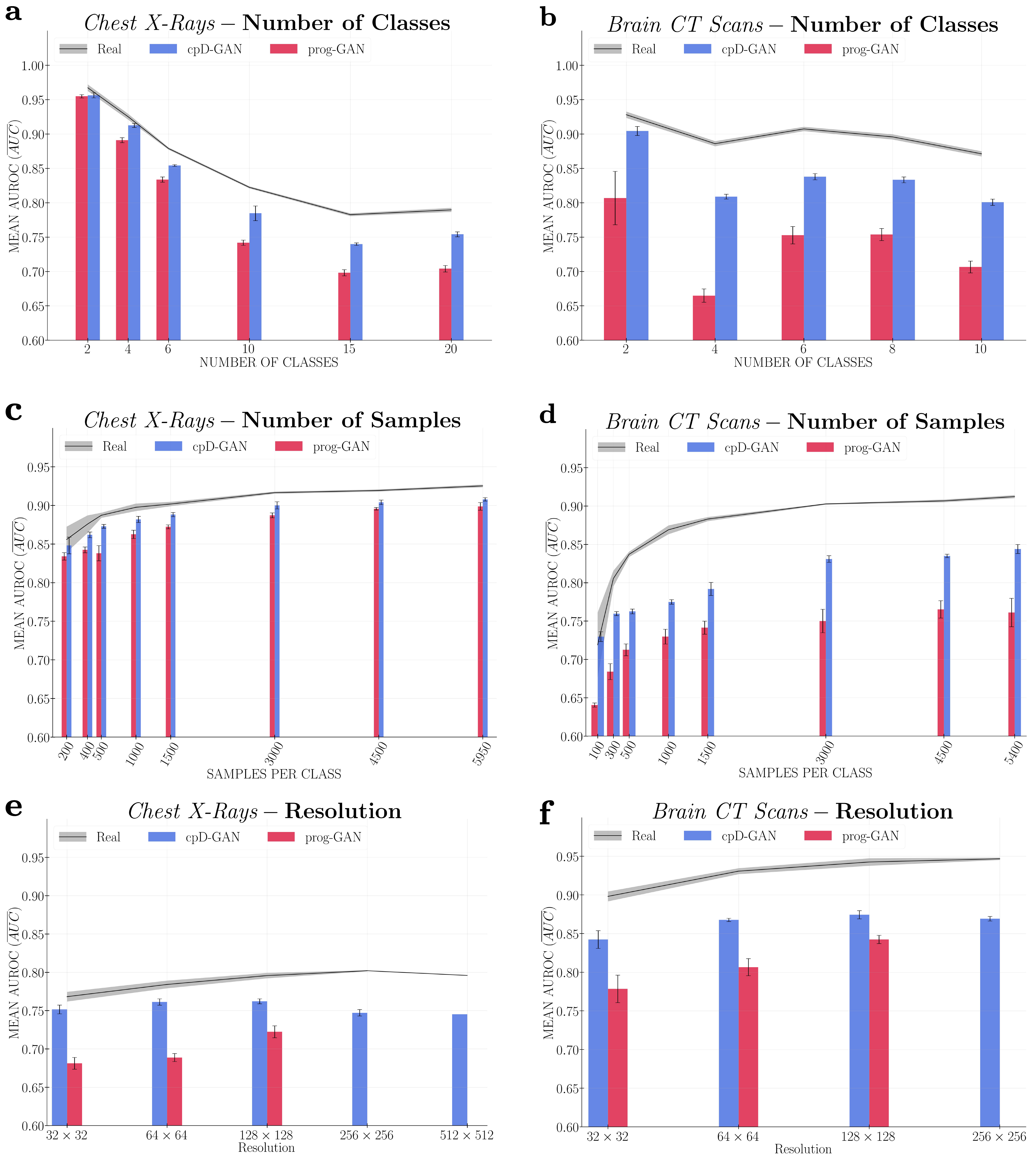}
\caption{\small{\textbf{Benchmark results.} In each figure, we show the mean area under the receiver operator characteristic curve score $(\overline{AUC})$ on a held-out test fold of real data. The $\overline{AUC}$ scores obtained after training classifiers on real data ($\overline{AUC}_{\rm real}$) are indicated by the black line with the shaded area representing the standard deviation across repeated experiments. The bar plots represent the $\overline{AUC}$ scores achieved after training classifiers on synthetic data ($\overline{AUC}_{\rm syn}$) generated by the cpD-GAN (blue) and prog-GAN (red), while the error bars indicate the standard deviation. The sub-figures show the changes in predictive performance observed when varying the number of classes (or label combinations) \textbf{a)} for chest radiographs and \textbf{b)} for brain computed tomography (CT) scans, the number of samples per class \textbf{c)} for chest radiographs and \textbf{d)} for brain CT scans, and the image resolution \textbf{e)} for chest radiographs and \textbf{f)} for brain CT scans. In \textbf{e)} and  \textbf{f)} only the cpD-GAN is evaluated at resolution levels above $128 \times 128$ pixels. In \textbf{e)} at $512 \times 512$ pixels we perform a single training run. Please see Table \ref{tab:1} for more details on the dataset composition for each benchmark setting.}}
\label{fig:2}
\end{figure}

\subsection{Benchmark findings} 
 
 We evaluated the model performance across three benchmark dimensions, detailed in Table \ref{tab:1}. First, we varied the number of unique binary label combinations (which we also refer to as number of classes) included in the dataset. Next, we fixed the present classes and assessed how changes in the number of samples for each group of findings impacted performance. While we evaluated the first two benchmark settings at a resolution of $32 \times 32$ pixels, we finally analysed how increasing the resolution to $64\times64$ and $128\times128$ pixels affected our scores.
 Due to the substantial computational demand at high spatial resolution, we only evaluate the cpD-GAN at $256 \times 256$ pixels for brain CT scans and $256 \times 256$ and $512 \times 512$ pixels for chest x-rays. We only performed changes across a single benchmark dimension at a time to ensure no confounding factors can impact training. 

\paragraph{Impact of number of classes.} The classification performance on both real and synthetic data increased when we lowered the number of unique present classes. We reason that the complexity of the predictive task decreases with fewer label combinations, resulting in higher $\overline{AUC}$ scores. However, as can be seen in Figure \ref{fig:2}a and \ref{fig:2}b, the differences in $\overline{AUC}_{\rm real}-\overline{AUC}_{\rm syn}$ scores also decreased when lowering the number of classes. For both datasets, the differences between the extreme cases ($20$ and $2$ classes for the chest x-rays and $10$ and $2$ classes for the brain CT scans) for the cpD-GAN were statistically significant ($\text{p-values}<0.0001$). The relative performance increase was even more pronounced for the prog-GAN. The trend of improvement in classifier performance when trained on synthetic data versus the performance when trained on real data shows that GAN models and the generated data quality disproportionately benefited from a smaller label space, thereby confirming the significance of the class conditioning methods. One crucial difference between the two evaluated GAN models is the improved label conditioning mechanism used with the cpD-GAN. The improved label conditioning was partially responsible for the lower overall scores and also explains why the prog-GAN had a more significant relative performance improvement on chest radiographs: Due to its inferior conditioning, the prog-GAN model benefited from a lower class number to a greater extent.

\paragraph{Impact of number of samples per class.} When we lowered the number of samples per label combination included in each dataset, the predictive performances obtained when training on real and synthetic data remained  similar until approximately $3,000$ samples per class. These results indicate that GAN model performance may be stable when the training data consists of at least $3,000$ samples per class. Between $1,500$ and $3,000$ samples, both the $\overline{AUC}_{\rm real}$ and $\overline{AUC}_{\rm syn}$ scores started to decrease substantially. However, we also observed a relative performance improvement, meaning decreasing $\overline{AUC}_{\rm real}-\overline{AUC}_{\rm syn}$ scores for the cpD-GAN, when moving towards low numbers of samples. This effect was particularly strong for the brain CT scans, where in the extreme setting of $100$ samples per class $\overline{AUC}_{\rm real}-\overline{AUC}_{\rm syn} \leq 0$. Despite the heightened variance in predictive performance, the difference in the scores between the extrema ($5,950$ and $200$ samples per class for the chest x-rays and $5,400$ and $100$ samples per class for the brain CT scans) was statistically significant ($ \text{p-values}<0.001$). The observed trend in performance in the low data regime indicates the growing effects of label overfitting during GAN training: Given a low number of samples, the variation within real images becomes too low, and the generative model may resort to encoding the class information in unrealistic ways. 

\begin{figure}[h!]
\center
\includegraphics[width=1.\linewidth]{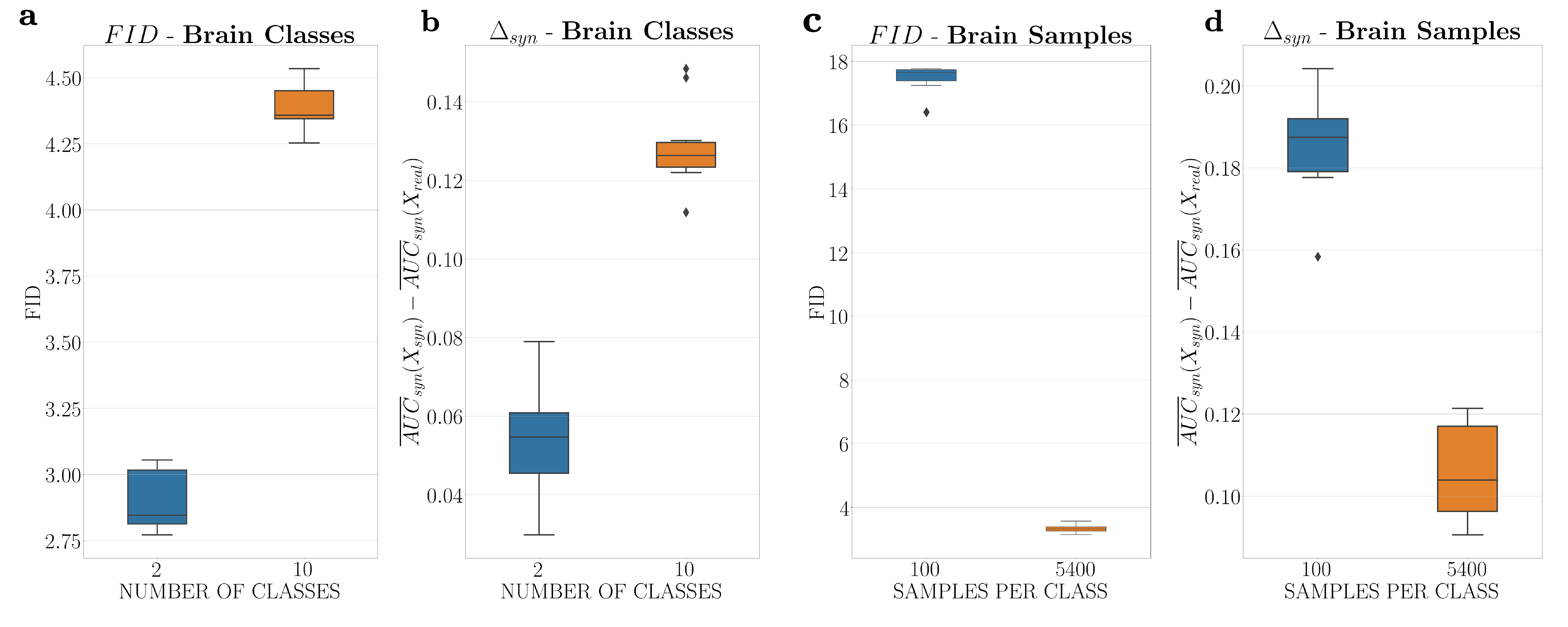}
\caption{\small{\textbf{Label overfitting analysis of the cpD-GAN on brain CT scans for the extrema settings from our benchmark:} Box plots showing the median, the interquartile range $(\text{IQR} = \text{Q}3-\text{Q}1)$, the minimum $(\text{Q}1-1.5\text{IQR})$, the maximum $(\text{Q}3+1.5\text{IQR})$ and outliers for \textbf{a)} FID scores between $10,000$ real and synthetic images after GAN convergence for $2$ and $10$ classes. \textbf{b)} $\Delta_{syn}$ scores, the difference between testing the predictive model on synthetic and real images after training on synthetics for $2$ and $10$ classes. \textbf{c)} FID scores between $10,000$ real and synthetic images after GAN convergence for $100$ and $5,400$ samples per class. \textbf{d)} $\Delta_{syn}$ scores, the difference between testing the predictive model on synthetic and real images after training on synthetics for $100$ and $5,400$ samples per class. Please see Table \ref{tab:1} for more details on the dataset composition for the extrema settings from our benchmark.}}
\label{fig:3}
\end{figure}

To further investigate our claim we analysed two more metrics for the extrema on the number of classes  ($2$ vs. $10$) and the number of samples per class ($100$ vs. $5,400$) benchmark for the brain CT scans. 1.) We looked at the distribution of FID scores for $10,000$ real and synthetic images after GAN training convergence. Low FID scores indicate synthetic images that are consistent and with an equivalent visual quality when compared to reals. The FID score is computed in a batch-wise fashion, which means that if the number of real training images is below $N=10,000$ we repeatedly iterate through the real dataset. While the labels that are fed into the GAN will likewise be repetitive, the generated synthetic images will differ due to different input noises. To have comparable FID scores, we need to keep $N=10,000$ constant, as the FID metric is biased with respect to the size of the sample set \cite{fidn_paper}. 2.) We analysed the difference in $\overline{AUC}$ scores when testing on synthetic and real images for classifiers trained on the synthetic dataset $\Delta_{syn} = \overline{AUC}_{\rm syn}(X_{syn}) -\overline{AUC}_{\rm syn}(X_{real})$. In the ideal scenario where our synthetic images have captured the real data distribution, we should observe $\Delta_{syn} \rightarrow 0$. If the GAN model faces label overfitting effects by generating images that encode the label information in unrealistic ways the $\Delta_{syn}$ scores should significantly increase. 

In Figure \ref{fig:3} we show box plots for the FID and $\Delta_{syn}$ scores for the extrema on the number of classes and number of samples per class benchmark for brain CT scans. In the number of classes settings we observed the expected behaviour: When including only $2$ classes, both the distribution of FID and $\Delta_{syn}$ scores was significantly lower compared to $10$ classes, which is in agreement with lower $\overline{AUC}_{\rm real}-\overline{AUC}_{\rm syn}$ scores from our benchmark, showing the improvement in GAN training due to a reduced label space. 

However, when we compared the settings for $100$ and $5,400$ samples per class we observed that in the low data regime we have significantly higher FID and $\Delta_{syn}$ scores. We, therefore, believe that the low $\overline{AUC}_{\rm real}-\overline{AUC}_{\rm syn}$ scores from our benchmark arose due to label overfitting effects. When entering the low data regime the variation in the GAN training data became too low resulting in synthetic images with lower visual quality (high FID scores). Because the GAN model was not able to generate realistic images, it started to overfit on the class information by unrealistic label encoding. The problem that generators are encouraged to produce images that are particularly easy for auxiliary classifiers to classify has been observed in the literature for several GAN conditioning mechanisms \cite{odena17a, miyato2018cgans}, and was one of the motivations for the conditional projection-based discriminator of our cpD-GAN \cite{miyato2018cgans}. It is an important finding that these effects can also occur for projection-based discriminators, when moving towards low data regimes. Analysing the exact way in which label information was encoded in the synthetic images and how this encoding generalized to testing on real data in our $\overline{AUC}_{\rm real}-\overline{AUC}_{\rm syn}$ score computation, remains open and an important direction for future work.

\paragraph{Impact of resolution.} When increasing the resolution from $32\times 32$ pixels to $128\times 128$ pixels, all $\overline{AUC}$ scores improved, as shown in Figure \ref{fig:2}e and \ref{fig:2}f. However, in terms of relative performance we observed a different behaviour for the two GAN models. For the prog-GAN, we observed a slight increase in  $\overline{AUC}_{\rm real}-\overline{AUC}_{\rm syn}$ scores at a resolution of $64\times 64$ pixels, with substantially lower scores at  $128 \times 128$ pixels. For the cpD-GAN, the predictive performance on real data increased disproportionately more, resulting in increased $\overline{AUC}_{\rm real}-\overline{AUC}_{\rm syn}$ scores on both datasets. Above $128 \times 128$ pixels, the $\overline{AUC}_{\rm real}-\overline{AUC}_{\rm syn}$ scores for the cpD-GAN deteriorate further. In general, GAN model training at higher resolutions is less stable and becomes more difficult due to the emergence of fine-scale details in the images. Moreover, the significant compute demand makes it more difficult to fine-tune the model hyper-parameters at these scales. While we conducted a large-scale hyper-parameter search for the high-resolution settings, that accounted for another $20,480$ GPU-hours on NVIDIA's Pascal P100, we could not find training parameters that improved from the previous settings. The negative effect on performance can be clearly seen for the cpD-GAN experiments at a resolution of $256 \times 256$ and $512 \times 512$ pixels on the chest x-rays. Compared to CT scans, radiographs are analysed at a higher spatial resolution, and the resulting lower synthetic data quality due to an increased training complexity is accurately detected in the benchmark evaluation. This is in accordance with the results from our reader study and the emergence of visual artifacts for the support devices class. Up until $128 \times 128$ pixels, we hypothesize that because the prog-GAN model is better fine-tuned on a more considerable number of resolution settings, it can achieve a relative performance improvement, compared to the cpD-GAN. Given an even greater amount of computational resources, it remains open whether the cpD-GAN can be fine-tuned to also benefit from resolutions above $256 \times 256$ pixels. Scaling the generation of synthetic medical images to even higher resolutions remains an area of active research (\cite{ quiros2020pathology, han2019breaking}), and is an important direction for future studies.

\begin{figure}[h!]
\center
\includegraphics[width=.8\linewidth]{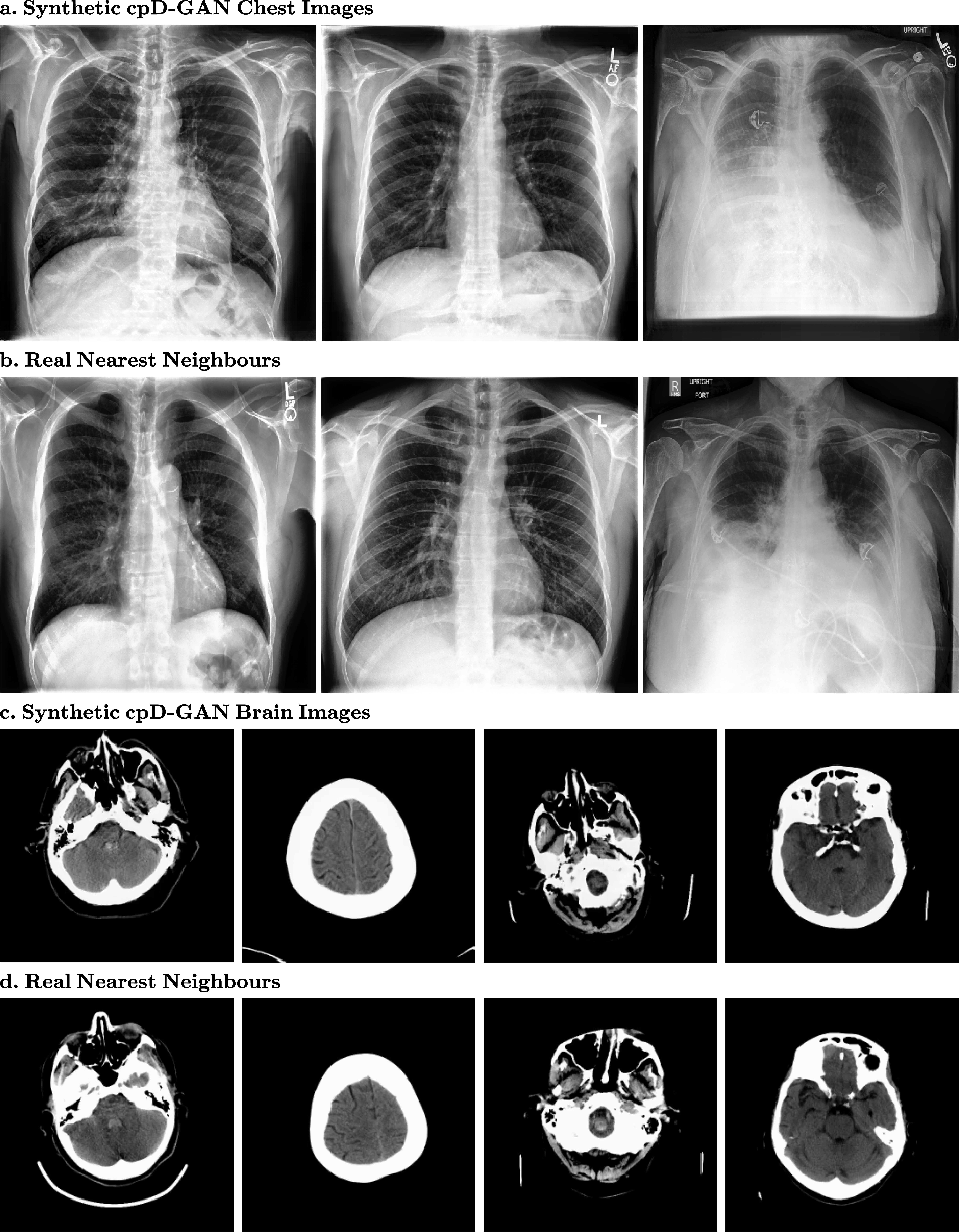}
\caption{\textbf{ Randomly sampled synthetic images generated by the cpD-GAN and real nearest neighbour images from the training.} \textbf{a)} Synthetic chest radiographs at $512 \times 512$ pixels. \textbf{b)} Nearest matching real images found in the chest radiograph training set. \textbf{c)} Synthetic brain computed tomography (CT) scans at $256 \times 256$ pixels. \textbf{d)} Nearest matching real images found in the brain CT training set.}  
\label{fig:4}
\end{figure}

\subsection{Further evaluation}

\begin{figure}[h!]
\center
\includegraphics[width=.6\linewidth]{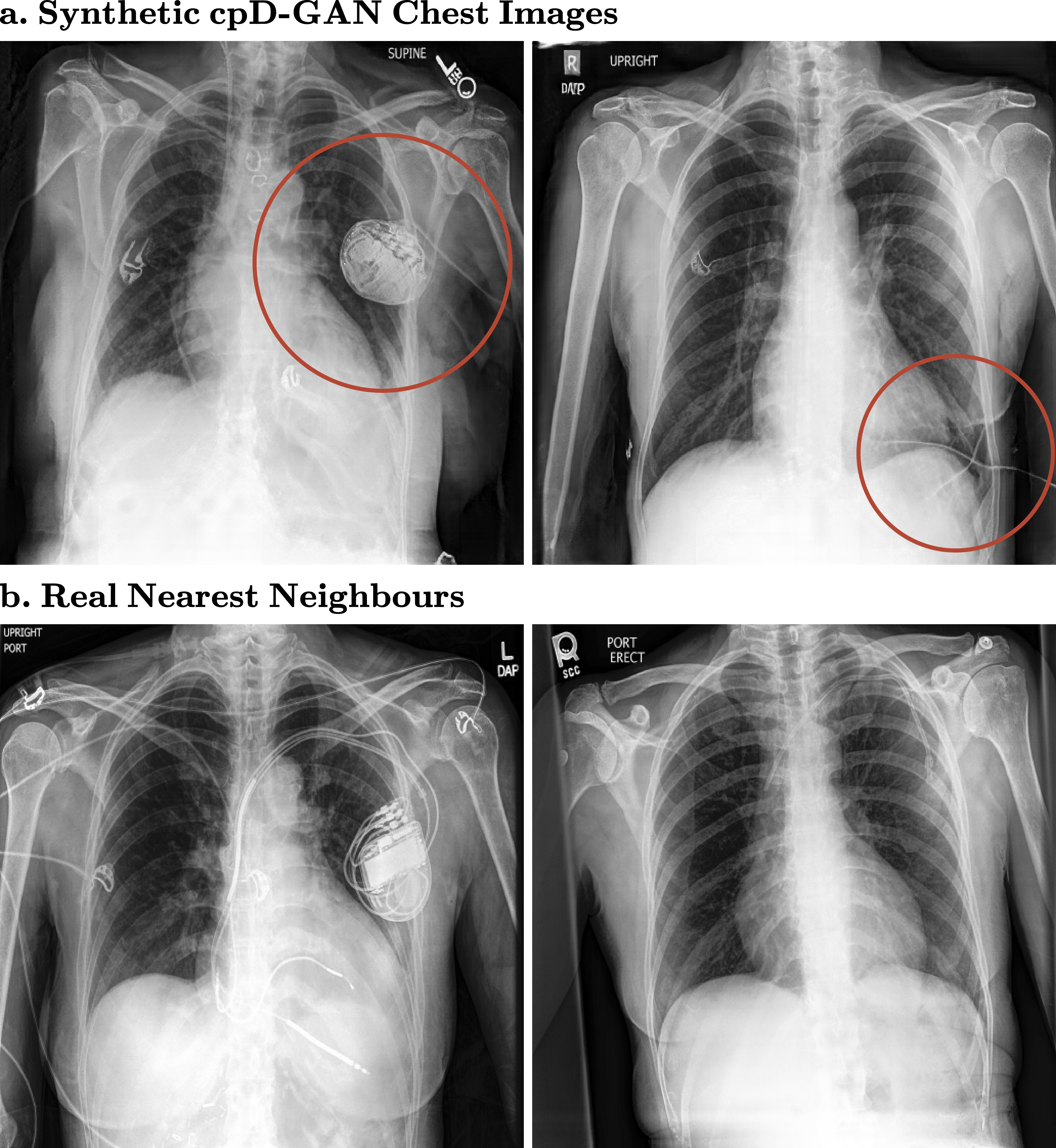}
\caption{\textbf{ a) Synthetic images with visual artifacts, generated by the cpD-GAN and b) real nearest neighbour images from the training dataset at $512 \times 512$ pixels.} The red circles surround unrealistic image neighbourhoods: Artifacts resulting from the support devices class.}
\label{fig:5}
\end{figure}

\begin{figure}[h!]
\center
\includegraphics[width=.8\linewidth]{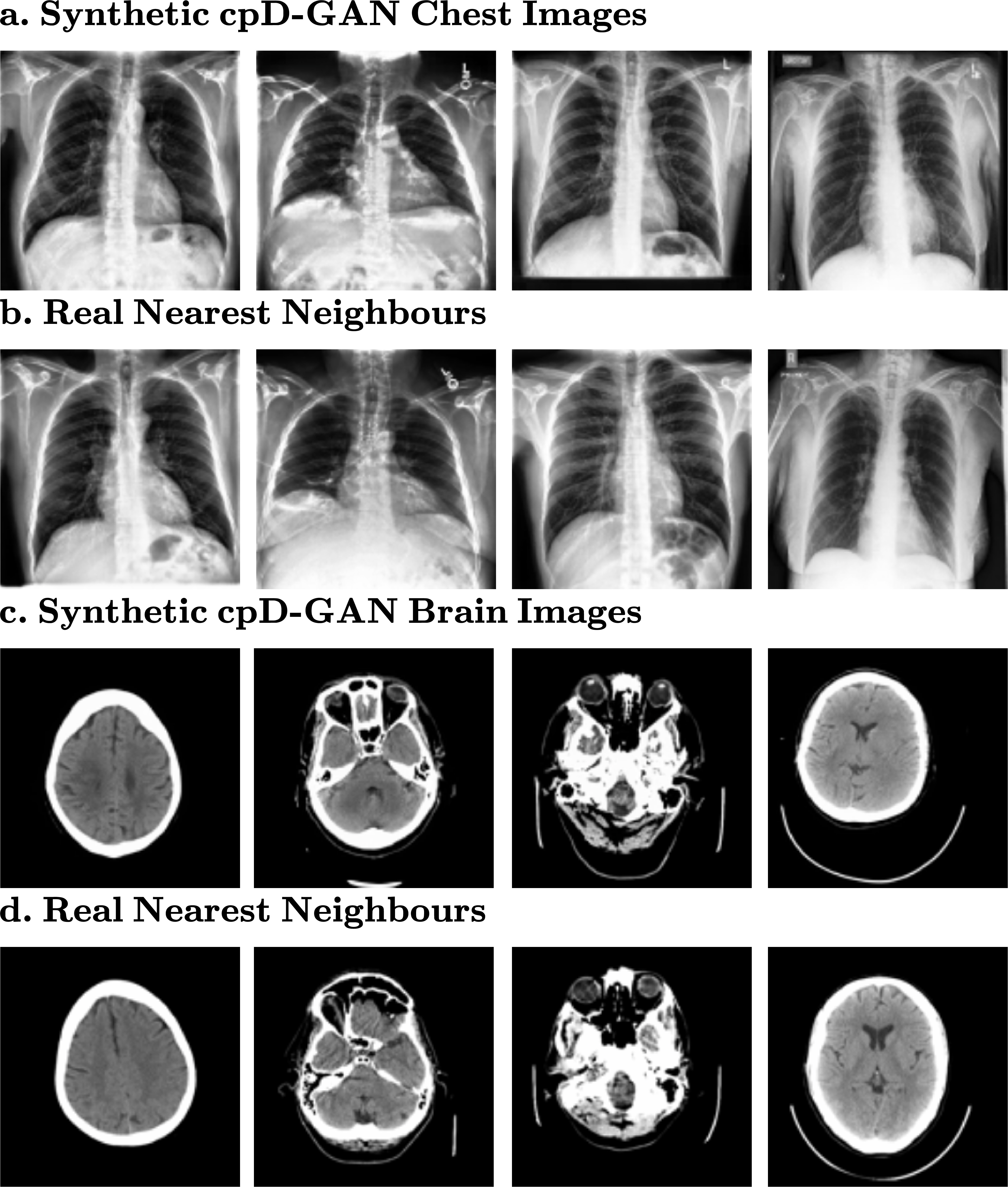}
\caption{\small{\textbf{More randomly sampled synthetic images from the cpD-GAN and nearest neighbours from all real training images at a resolution of $128 \times 128$ pixels.} \textbf{a)} Synthetic chest radiographs. \textbf{b)} Nearest matching real images found in the chest radiograph training set. \textbf{c)} Synthetic brain computed tomography (CT) scans. \textbf{d)} Nearest matching real images found in the brain CT training set.}}
\label{fig:s4}
\end{figure}

\begin{figure}[h!]
\center
\includegraphics[width=1.\linewidth]{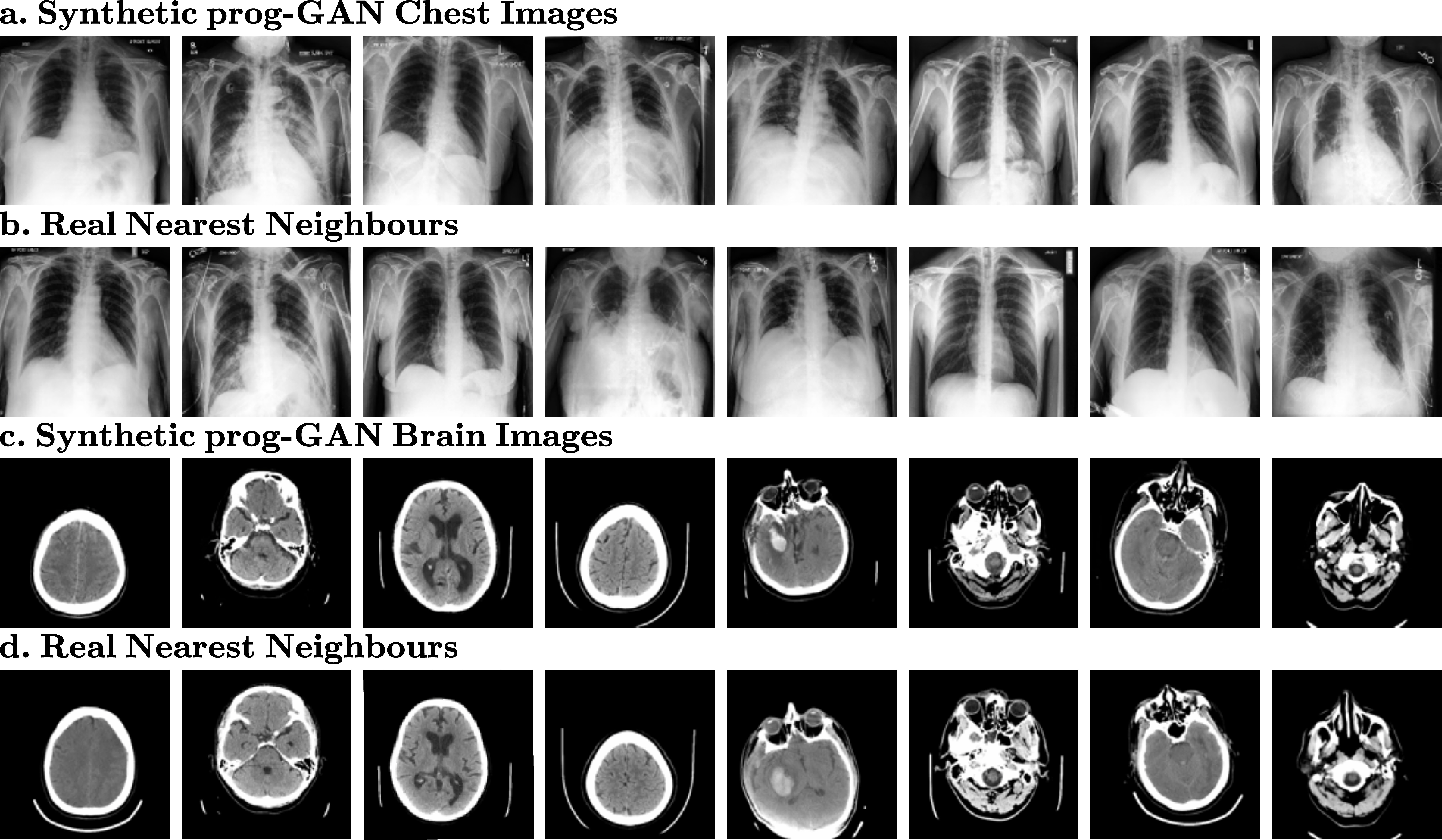}
\caption{\small{\textbf{Randomly sampled synthetic images from the prog-GAN and nearest neighbours from all real training images at resolution of $128 \times 128$ pixels.} \textbf{a)} Synthetic chest radiographs. \textbf{b)} Nearest matching real images found in the chest radiograph training set. \textbf{c)} Synthetic brain computed tomography (CT) scans. \textbf{d)} Nearest matching real images found in the brain CT training set.}}
\label{fig:s6}
\end{figure}

\textbf{Image quality and privacy.} In Figure \ref{fig:4}, we show randomly sampled synthetic example images from the cpD-GAN at a resolution of $512 \times 512$ pixels for chest x-rays and $256 \times 256$ pixels for brain CT scans. Below each synthetic image, we show the most similar real image (nearest neighbour) out of the entire training dataset. For the brain CT scans, there appears to be little noticeable difference in visual quality between the real and synthetic images, which is in agreement with a close-to-random classification accuracy of trained radiologists in the reader study. In contrast, while the chest x-rays also have a high visual quality, there are differences between reals and synthetics, which become more apparent in the form of artifacts for one particular class. The cpD-GAN failed to realistically generate tubes and other support devices, such as pacemakers or defibrillators, as shown in Figure \ref{fig:5}. These devices deviate strongly in their visual appearance when compared to the physiological chest outlining and were not accurately learned by the generative model. Crucially, our benchmark successfully captured the drops in visual quality, as indicated by higher $\overline{AUC}_{\rm real}-\overline{AUC}_{\rm syn}$ scores. Given the frequent presence of support devices, the GAN should ideally also learn their data distribution. Nonetheless, the cpD-GAN captured the distribution of the physiological anatomic chest structure and related radiology findings which is more important with respect to diagnosis and clinical practice than external objects. At a spatial resolution of $128 \times 128$ pixels the visual differences are not yet apparent (see Figure \ref{fig:s4}) which is in accordance with a better benchmark performance. Even in settings where the GAN failed to learn the data distribution of a specific class, it did not start to copy training images: By comparing synthetics and nearest matching neighbours we demonstrate that the cpD-GAN model did not simply memorize training data, and is therefore likely to preserve private, potentially sensitive information. For more high-resolution example images from the cpD-GAN and the respective nearest neighbours see Figure \ref{fig:s3} in the supplementary materials. From a visual inspection, the quality of the images generated by the prog-GAN appear to be only marginally worse than those generated by the cpD-GAN at $128 \times 128$ pixels (Figure \ref{fig:s6}).

\newpage 

\begin{figure}[h!]
\center
\includegraphics[width=.8\linewidth]{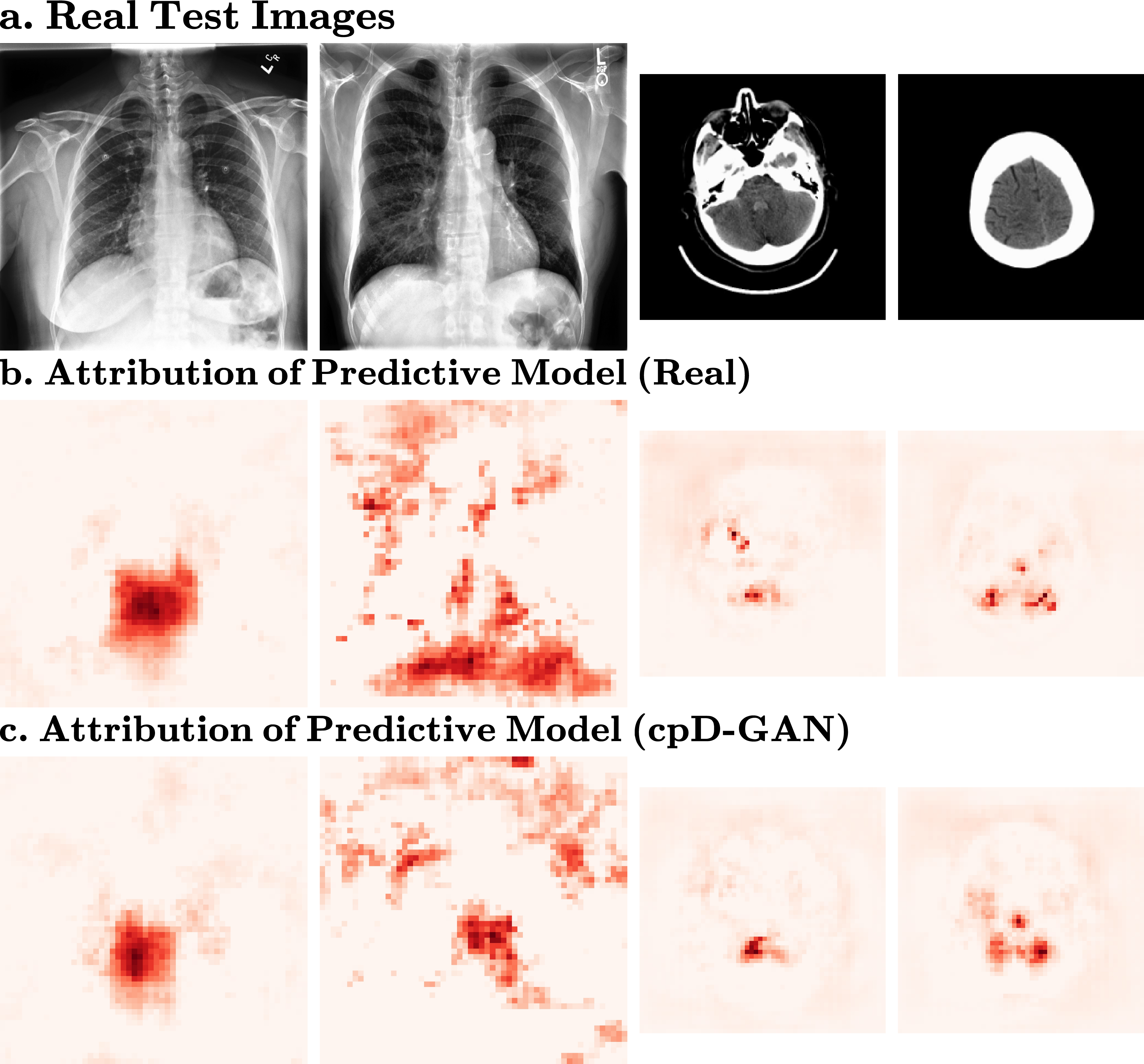}
\caption{\textbf{Feature importance of predictive models.} Deeper red colour indicates regions that have a larger causal contribution to the label prediction. \textbf{a)} Real images (nearest neighbours from Figure \ref{fig:4}, $512 \times 512$ pixels for chest x-rays, $256 \times 256$ pixels for brain CT scans). From left to right: (1) Chest x-ray with no finding. (2) Chest x-ray with cardiomegaly, atelectasis and support device (3) Brain scan with no finding. (4) Brain scan with intraventricular hemorrhage. \textbf{b)} Feature importance of predictive model trained on real data. \textbf{c)} Feature importance of predictive model trained on synthetic data generated by the cpD-GAN. \textbf{d)} Feature importance of predictive model trained on synthetic data generated by the prog-GAN.}
\label{fig:6}
\end{figure}

\begin{figure}
\center
\includegraphics[width=.6\linewidth]{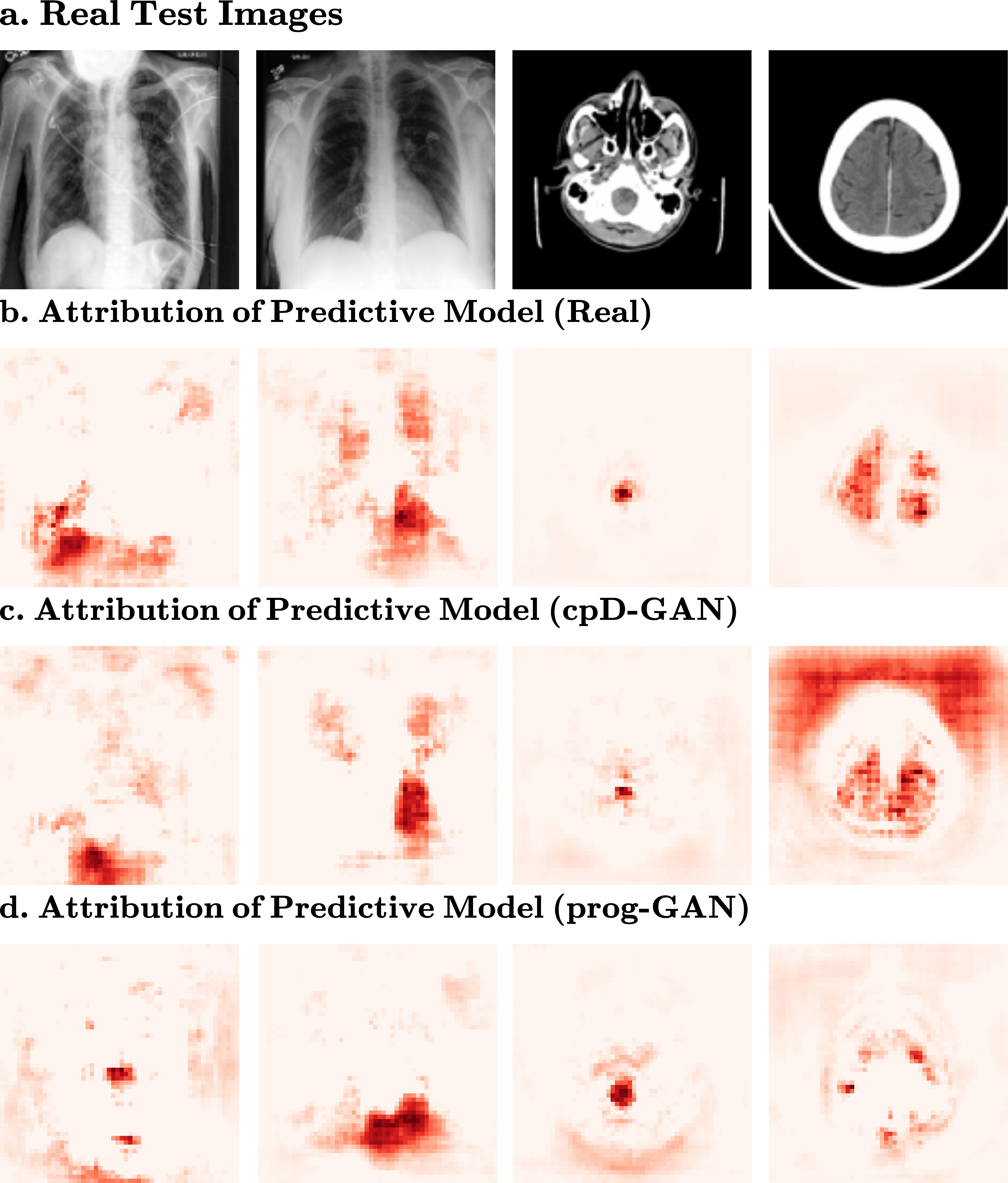}
\caption{\textbf{Feature importance of predictive models.} Deeper red colour indicates regions that have a larger causal contribution to the label prediction. \textbf{a)} Real test images (nearest neighbours from Figure \ref{fig:4}) at $128\times 128$ resolution. From left to right: (1) Chest x-ray with support device, lung opacity, pneumonia and atelectasis. (2) Chest x-ray with cardiomegaly and edema. (3) Brain scan with subarachnoid hemorrhage. (4) Brain scan with subdural hemorrhage. \textbf{b)} Feature importance of predictive model trained on real data. \textbf{c)} Feature importance of predictive model trained on synthetic data generated by the cpD-GAN. \textbf{d)} Feature importance of predictive model trained on synthetic data generated by the prog-GAN.}
\label{fig:7}
\end{figure}

\begin{figure}[h!]
\center
\includegraphics[width=.8\linewidth]{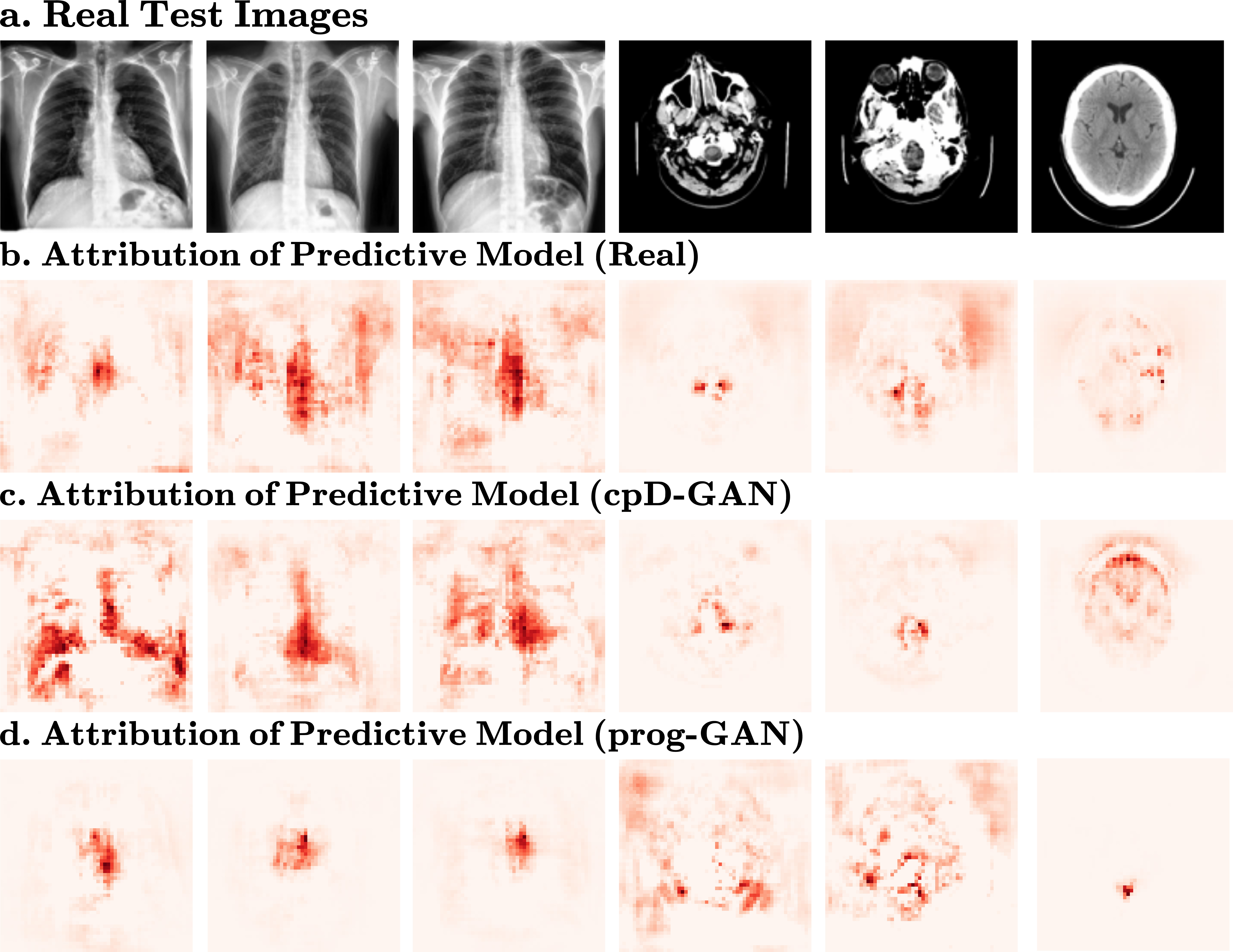}
\caption{\small{\textbf{More feature importance maps of predictive models.} Deeper red colour indicates regions that have a larger causal contribution to the label prediction. \textbf{a)} Real test images (nearest neighbours from Figure \ref{fig:4}) at $128\times 128$ resolution. All displayed images are without any clinical finding. \textbf{b)} Feature importance of predictive model trained on real data. \textbf{c)} Feature importance of predictive model trained on synthetic data generated by the cpD-GAN. \textbf{d)} Feature importance of predictive model trained on synthetic data generated by the prog-GAN.}}
\label{fig:s5}
\end{figure}

\paragraph{Feature importance.} To gain more interpretability, we analysed the feature importance at a resolution of $512 \times 512$ pixels for chest x-rays and $256 \times 256$ for brain CT scans from the cpD-GAN, as well as at $128 \times 128$ pixels for both datasets and models. In each setting, we estimate the feature importance by successively masking out pixel regions and computing the increased loss \cite{schwab2019cxplain}. Similar attribution maps indicate that the pixel neighbourhoods in real and synthetic images have a similar causal loss contribution, leading to equivalent predictive models. This suggests that the local image neighbourhoods are consistent in style and texture. While the datasets only allow for classification tasks, we can extrapolate from this analysis that our synthetic images might also perform similarly on problems such as object detection or segmentation, which require global and local image consistency. Instead of randomly sampling real test images, we chose the real nearest neighbours from Figure \ref{fig:4}. Figure \ref{fig:6} shows the nearest neighbours, the corresponding attribution maps of the predictive model trained on reals, and the attribution maps for those trained on the synthetic images generated by the cpD-GAN at high resolution. In Figure \ref{fig:7}, the same analysis is performed at a resolution of $128 \times 128$ for both the cpD-GAN and the prog-GAN (for more examples see Figure \ref{fig:s5}). The observed results support the hypothesis that the predictive models trained on synthetic data from the cpD-GAN assign importance to similar image features as those trained on real data. In accordance with our benchmark results, the feature maps of the cpD-GAN appear more similar to those of real classifiers for a resolution of $128 \times 128$ pixels (and $256 \times 256$ for brain CT scans), while the differences are greater at $512 \times 512$ pixels for chest x-rays. Moreover, in line with the higher $\overline{AUC}_{\rm real}-\overline{AUC}_{\rm syn}$ scores for the prog-GAN synthetics, the attribution maps also appear to be visually more dissimilar from those assigned by classifiers trained on real data. We note that none of the feature importance maps were identical, which we expected given that the observed difference in predictive performance between classifiers trained on real and synthetic data was greater than zero at those settings. Please refer to Section \ref{sec:feat_imp} for more details on the attribution map computation.

\paragraph{Reader study.} We additionally conducted a reader study in which we asked trained radiologists to label a mixed set of $100$ images as real or synthetic (generated by the cpD-GAN) at a resolution of $512 \times 512$ pixels for chest x-rays, $256 \times 256$ for brain CT scans and $128 \times 128$ pixels for both data modalities. In terms of results, we found that radiologists were unable to achieve a higher accuracy than a classifier assigning labels at random with an expected accuracy of 50\% ($p < 0.05$ for chest radiographs, $p < 0.01$ for brain CT scans) in both $128 \times 128$ pixels settings. This is in agreement with relatively low $\overline{AUC}_{\rm real}-\overline{AUC}_{\rm syn}$ scores from our benchmark evaluation. Radiologists were able to differentiate between real and synthetic brain CT scans at $256 \times 256$ pixels with an accuracy of $0.532 \ \pm 0.126$, which is still close to random, indicating a consistent visual quality and the absence of visual artifacts. In line with higher $\overline{AUC}_{\rm real}-\overline{AUC}_{\rm syn}$ scores and the emergence of visual artifacts, radiologists achieved an $0.710 \ \pm 0.148$ accuracy when classifying the chest x-rays at $512 \times 512$ pixels. The presented results indicate that at lower spatial resolution trained clinicians cannot discriminate between real and synthetic images, which further substantiates that both the general quality and label information in the synthetic images are realistic. In agreement with our other results, the classification accuracy of radiologists improves with an increasing spatial resolution due to fine-scaled image details and the emergence of visual artifacts, especially for the x-rays. Please refer to Section \ref{sec:met_rs} for details on the set-up and to Section \ref{sec:met_conf} for a presentation of the detailed results of the conducted reader study.

\section{Discussion}
In this study, we benchmarked the generation of synthetic medical image data to closely mimic the distribution-level statistical properties of a real source dataset. To do so, we evaluated two state-of-the-art GAN models, prog-GAN and cpD-GAN, on two real-world medical image corpora consisting of chest radiographs and brain CTs, respectively. We compared the difference in performance on real test data between a predictive model trained only on real or only synthetic images. As part of the conducted benchmark evaluation, we analysed the effects of changes in the number of label combinations, samples per class, and resolution. The presented results offer valuable guidelines for synthesising medical imaging datasets in practice. In addition, we analysed the difference in causal contributions of predictive models when trained on either the real or synthetic dataset and investigated the privacy-preservation in our generated medical images by comparing them to the most closely matching real training images. We found that synthetic medical images generated by the cpD-GAN enabled training of classifiers that closely matched the performance of classifiers trained on real data. Finally, we conducted a large-scale reader study in which we found that trained radiologists could not discriminate better than random between real and synthetic images, generated by the cpD-GAN, for both datasets at a resolution of $128 \times 128$ pixels. Our benchmark evaluation and detailed analysis of the synthetic images also shows limitations to medical data generation: Due to an increasing training complexity at higher spatial resolution levels, the GAN models fail to accurately learn the data distribution for the support devices class on chest x-rays. This leads to a decreased benchmark performance and synthetic images with a lower visual quality when compared to the real data.

\paragraph{Generalisation ability.} We determined that both GAN models are stable across all benchmark dimensions up to $128 \times 128$ pixels, meaning that we did not observe anomalous $\overline{AUC}_{\rm real}-\overline{AUC}_{\rm syn}$ scores. While some GAN models that we trained were not robust across the experiments (see Section \ref{sec:biggan}), the prog-GAN and cpD-GAN did not collapse at any setting or choice of random initialisation. There were no hyper-parameter changes for the varying experiments or across the two datasets. Only when increasing the spatial resolution, we added the necessary convolutional blocks to both models. While the $\overline{AUC}_{\rm real}-\overline{AUC}_{\rm syn}$ scores above  $128 \times 128$ pixels further increased, there was no training collapse. The visual quality of brain CT scans remained high at $256 \times 256$ pixels. For the x-rays we observed that at a resolution of $512 \times 512$ pixels, generated objects from the support devices class suffered from a lower visual quality, resulting in an improved radiologist classification. Crucially, the reduced performance in our benchmark findings supported our detailed analysis of the synthetic images. Also, the cpD-GAN did not fail to capture the distribution of the physiological anatomic structure of the chest, which might be more important for clinical practice. The observed results indicate that the presented GAN pipeline is robust to changes in the dataset-, and data modality, and that it may generate high quality synthetic medical images across various conditions with the desired statistical similarity compared to the training cohort. For synthetic data generation to work reliably in practice, the convergence of the generative models and the quality of the generated images must be robust across different cohorts where the number of available samples or classes might deviate. While we focused on $2$ datasets, our generative methods and evaluation protocols can be easily extended to different settings and are not limited to the chest radiographs and brain CT scans. We believe that our findings show that sharing synthetic medical imaging datasets may be an attractive and privacy-preserving alternative to sharing real patient-level data in certain settings, thereby providing a technical solution to the pervasive issue of data sharing in medicine \cite{lo2015}. 

\paragraph{Practical guidelines.} The predictive performance obtained when training on synthetic data improved when reducing the number of classes present in the dataset. The impact of a reduced label spaces suggests that researchers should, in practice, choose datasets for GAN model development that have a manageable number of unique label combinations. Even though rare findings may be particularly interesting from a clinical perspective, they should be excluded from training when maximum performance is required, since it is currently impossible to give any guarantees for consistent quality in low sample numbers containing rare findings. Moreover, the samples per class benchmark indicates that the GAN models might overfit on rare classes by encoding label information in unrealistic ways within the synthetic images. This can be very problematic as it can lead to predictions that are based on features not present in the real data distribution when trained on synthetics. The most critical performance improvement between the prog-GAN and cpD-GAN resulted from a revised label conditioning mechanism, rooted in a probabilistic framework \cite{miyato2018cgans}. Therefore, the impact of the class conditioning mechanism on the predictive performance of derived classifiers suggests that research on the conditioning mechanism of GAN models may lead to further improvements in image quality. In the chest radiograph benchmark, we found that the total number of samples in the training dataset can be lowered significantly (to approximately $9,000$), at a number of samples per class of around $3,000$ without any relative performance drop. However, if low-frequency classes are included and the total number of samples is reduced too much, GAN label overfitting is likely to occur.

 In terms of predictive performance in relation to different image resolutions, we found that $\overline{AUC}_{\rm syn}$ scores for both models improved when moving up to $128 \times 128$ pixels. However, we also observed different behaviours in the different GAN models in terms of relative performance compared to real data when adjusting the image resolution. For the prog-GAN, relative predictive performance increased at a resolution of $128 \times 128$ pixels compared to lower levels, which indicates that, once a GAN model has been fine-tuned, it can benefit from the emergence of details at a higher spatial resolution. We note that the prog-GAN hyper-parameter settings were taken from the official implementation \cite{karras2018progressive}, which has been well adjusted to a number of datasets. While our cpD-GAN model outperformed the prog-GAN at all evaluated resolutions, its own $\overline{AUC}_{\rm real}-\overline{AUC}_{\rm syn}$ scores increased when moving up from $32 \times 32$ pixels. The dominance of training instabilities increased further when moving towards $256 \times 256$ and $512 \times 512$ pixels for chest x-rays. This behaviour is not unexpected as stable training becomes more difficult when the discriminator has access to a richer set of features to distinguish real and synthetic data. In line with our benchmark performance decrease and reader study results, we observed that at $512 \times 512$ pixels, unstable training resulted in cpD-GAN models that did not capture the real data distribution for the support devices class of chest x-rays. The cpD-GAN was not able generate consistent and high-quality objects that are not part of the physiological chest outlining, such as tubes, pacemakers, or defibrillators. While it is more important to accurately capture the data distribution of the physiological anatomic chest structure and related radiology findings, the GAN may ideally also generate visually appealing external objects. It remains open whether more hyper-parameter fine-tuning, a refined model architecture, or different training strategies, could improve inconsistencies in the synthetic images. In practice, researcher need to carefully consider at what spatial resolution fine-scaled details emerge that differ significantly from other image parts. Importantly, our benchmark evaluation captured these inconsistencies and robustly detected visual artifacts. For the brain CT scans we did not observe a lower image quality at a resolution of $256 \times 256$ pixels. Accordingly, trained radiologists performed only marginally better than random at differentiating between reals and synthetics.   
In the presented reader study, we found that at $128 \times 128$ pixels, the accuracy distribution derived from the real and synthetic labels set by radiologists was not better than that of a random classifier with a mean accuracy ($\overline{acc}$) of $50\%$, to a statistically significant extent. The fact that trained clinicians were unable to discriminate between the real and synthetic medical imaging datasets indicates that the generated images had a realistic visual appearance and label information was included in a qualitatively reasonable manner. The results of the presented reader study further support the findings presented in the conducted experimental benchmark evaluation and show that the cohort level information of medical imaging data can be shared without relying on patient-level data.

We have shown that, under the right conditions, sharing synthetic medical imaging datasets may be a viable alternative to real data sharing. However, the presented results also show that there is a measurable gap in quality and predictive performance between synthetic and real medical imaging data, especially when moving to high-resolution. Across all benchmark settings, we observed that only in the extreme label overfitting case $\overline{AUC}_{\rm real}-\overline{AUC}_{\rm syn} \leq 0$, meaning that in all other experiments, there was a reduced performance when training on the generated images. While this difference was relatively small for the cpD-GAN across the chest radiographs, it was more pronounced on the brain CT scans. From our causal contribution investigation in Figure \ref{fig:5} and \ref{fig:6}, we found that while the real and cpD-GAN predictive models attributed similar regions with high feature importance, they were not identical. In line with our other findings the differences appeared to increase when moving towards high resolutions. In the ideal case, both assigned feature importance and predictive performance would be identical when replacing real data with synthetic data. Even so, our benchmark results demonstrate that the goal of learning the real data distribution for medical images is realistic and feasible.

\paragraph{Limitations.}

In this study, we analysed synthetic chest radiographs at an overall low resolution compared to clinical practice. Even at $512 \times 512$ pixels, the images might lack important details for an accurate clinical diagnosis. A comparison of low-resolution images is more acceptable for evaluating the predictive models as most deep learning systems down-sample medical images to reduce the computational requirements. Since training generative models is even more computationally demanding, the lack of unlimited resources represents a major bottleneck for further up-scaling. However, our benchmark results indicate that state-of-the-art GAN models already have difficulties in accurately modeling the support devices class at the analysed spatial resolution. While GAN models such as \cite{Karras2019AnalyzingAI, brock2018large} and other generative approaches such as \cite{vahdat2020nvae, dhariwal2021diffusion} work at high-resolution levels, our observed drop in performance indicates that within the medical imaging domain, more model and training improvements are necessary to ensure that the full data distribution is learned in these regimes. Therefore, future work should focus on the current limitations shown in our benchmark evaluation, before scaling GAN training further.

The resolution at which we generate brain CT scans is only marginally smaller than the maximum resolution of the dataset and generally more acceptable when compared to medical practice. However, clinicians analyse 3D computed tomography scans at different intensity windows. Here, we were limited by the RSNA Intracranial Hemorrhage dataset, which consists of pre-sliced scans with only the soft-tissue window. To analyse a variety of different benchmark settings, we required a certain number of samples that rarely exist in open-source medical imaging datasets. Moreover, reliable synthetic medical data generation is currently limited to 2D settings as it becomes substantially more complex, both in terms of required computational resources and algorithmic challenges, to model 3D structures. 

Similarly, the lack of large-scale medical imaging corpora for object detection or segmentation, limited our ability to conduct experiments with downstream tasks, other than classification. Our results from the feature importance analysis and reader study suggest both local and global image consistency up to intermediate resolution levels. This indicates that our benchmark results may also hold for segmentation or detection performance. Nevertheless, it is an important direction for future work, once newly published medical imaging datasets allow for a similar large-scale benchmark evaluation on the aforementioned tasks. 

Finally, the presented study does not provide any mathematical guarantees for the privacy of the synthetic data. We found settings in which privacy would likely be breached in practice, which can be an important guideline, but a more formal analysis in terms of differential privacy may in the future further elucidate the degree to which generative modeling preserves individual patient-level information \cite{dwork2014algorithmic}. In Figure \ref{fig:4}, we demonstrate that there were considerable differences between the generated images and the most closely matching nearest neighbour images from the training data, which may indicate that the GAN models learn the actual data distribution and do not merely memorize the training set. However, a retrospective analysis may not always be feasible, and more formal privacy guarantees regarding the model and training may be needed in some real-world use cases. Through the use of stochastic gradient descent, all of our GAN models have some level of intrinsic privacy \cite{hyland2019intrinsic}, but it remains an area of active research to quantify how strong these privacy guarantees are. While there remain open questions for further research, our results indicate that synthetic data sharing may
in the future become an attractive and privacy-preserving alternative to sharing real patient-level data in the right settings.

\section{Data availability}

Both datasets used in our study are publicly available and free to download for any registered user. The CheXpert chest radiograph dataset \cite{irvin2019chexpert} can be accessed at \href{https://stanfordmlgroup.github.io/competitions/chexpert/}{https://stanfordmlgroup.github.io/competitions/chexpert/} and the RSNA Intracranial Hemorrhage dataset \cite{flanders2020construction} is available at \href{https://www.kaggle.com/c/rsna-intracranial-hemorrhage-detection}{https://www.kaggle.com/c/rsna-intracranial-hemorrhage-detection}. 

\section{Code availability}

To reproduce our benchmark results, please see our code repository under  \href{https://github.com/AugustDS/synthetic-medical-benchmark}{https://github.com/AugustDS/synthetic-medical-benchmark} (MIT license). 

\clearpage

\bibliography{main}
\bibliographystyle{unsrtnat}

\newpage

\beginsupplement
\section{Methods}
\label{sec:supp}

\subsection{Datasets and pre-processing} 

The CheXpert dataset consists of $224,316$ chest radiographs of $65,240$ patients, collected from radiographic examinations of the chest at the Stanford Hospital, between October 2002 and July 2017 \cite{irvin2019chexpert}. In the dataset study, an automatic labeling tool was used to identify and classify the certainty of the presence of $14$ observations from the radiology report. We turned uncertain labels into positives, to make use of all data, resulting in a binary multi-label dataset, where a large number of label combinations can co-occur. 

The RSNA Intracranial Hemorrhage Dataset is composed of computed tomography studies supplied by four research institutions and labeled with the help of The American Society of Neuroradiology \cite{flanders2020construction}. It consists of $752,803$ CT scan slices of the head from $18,938$ unique patients and the corresponding probabilities for the presence of $5$ different hemorrhage types and the no finding label. For consistency, we turned any probability $p_{y_i}>0$ into a positive label $y_i=1$ and else $y_i=0$, also resulting in a binary multi-label dataset. Since $644,874 \ (85.7\%)$ of CT scans are without any intracranial hemorrhage, we undersampled the no-finding class, resulting in a balanced dataset where at least $50\%$ of images show some form of hemorrhage. 

We randomly split the entire patient cohort into training ($80 \%$), validation ($10 \%$), and test folds ($10 \%$) within strata of radiology findings for each dataset. We excluded chest x-rays of classes with fewer then $256$ samples, resulting in $117,168$ train images ($44,153$ patients), $15,318$ validation images ($5,519$ patients) and $14,687$ test images ($5,520$ patients). For the hemorrhage dataset we removed label combinations below a frequency of $100$,  resulting in $173,271$ train images ($15,133$ patients), $22,095$ validation images ($1,892$ patients), and $20,500$ test images ($1,892$ patients). 

We developed the resolution benchmark for both datasets on the aforementioned setting. For the class benchmarking, we gradually reduced the number of clinical finding combinations present in the dataset, while keeping the total number of training images constant via over-sampling. When benchmarking the effect of samples per clinical finding, we fixed the number of classes and gradually decreased each class's frequency. Table \ref{tab:1} gives a complete summary of all dataset settings, the entire set of labels, the size of training, validation and test sets, and information on remaining labels and samples per class. Each summary refers to the real training, validation and testing dataset. The exact labels from the real settings (combined with random normal noise for variation) are used to generate equivalent synthetic training, validation and testing datasets, before developing and comparing the predictive models.

\clearpage

\begin{table}[h!]
\centering
\scalebox{0.65}{\begin{tabular}{|p{4.8cm}|l|l|l|l|l|l|l|}
\hline
 Dataset Information & Benchmark & Resolution & $m_{labels}$ & $m_{label \ comb}$ & $n_{tr}$ & $n_{val/te}$ & $n_{per \ label \ comb}$   \\ \hline

 \textbf{Chest Radiographs} & Classes & $32 \times 32$ & $9$ & $20$ & $29000$ & $3800$ & $1450$  \\

   &  & & $8$ & $15$ & $24000$ & $2850$ & $1600$  \\
  
  \textbf{Data Pool:} &  & & $5$ & $10$ & $20000$ & $1900$ & $2000$  \\
  
  $n_{tr} \ (n_{pat}) = 117168 \ (44153)$ &  & & $5$ & $6$ & $13800$ & $1140$ & $2300$ \\
  
  $n_{vl} \ (n_{pat}) = 15318 \phantom{8} \ (5519)$&  & & $5$ & $4$ & $15600$ & $760$ & $3900$ \\
  
  $n_{te} \ (n_{pat}) = 14687 \phantom{8} \ (5520)$ &  & & $4$ & $2$ & $12600$ & $380$ & $6300$ \\ 
   
   \cline{2-8}
   
   & Samples & $32 \times 32$ & $4$ & $3$ & $17850$ & $2250$ & $5950$ \\

   \textbf{All Labels:} & & & $4$ & $3$ & $13500$ & $2250$ & $4500$  \\
  
   Enlarged Cardiomediastinum, &  & & $4$ & $3$ & $9000$ & $2250$ & $3000$  \\
  
   Cardiomegaly, Lung Opacity, &  & & $4$ & $3$ & $4500$ & $2250$ & $1500$  \\
  
   Lung Lesion, Edema, &  & & $4$ & $3$ & $3000$ & $2250$ & $1000$  \\
  
   Consolidation, Pneumonia, &  & & $4$ & $3$ & $1500$ & $2250$ & $500$ \\
   
   Atelectasis, Pneumothorax, &  & & $4$ & $3$ & $1200$ & $2250$ & $400$  \\
   
   Pleural Effusion, Pleural Other, &  & & $4$ & $3$ & $600$ & $2250$ & $200$  \\

   \cline{2-8}
   
   Fracture, Support Device, & Resolution & $32 \times 32$& $14$ & $138$ & $117168$ & $4000$ & $256-7586$  \\
   
   No Finding& (pixels)  & $64 \times 64$& $14$ & $138$ & $117168$ & $4000$ & $256-7586$  \\
   
   &  & $128 \times 128$& $14$ & $138$ & $117168$ & $4000$ & $256-7586$  \\
     
   \hline
   \hline
   
   \textbf{Brain Hemorrhage CTs} & Classes & $32 \times 32$ & $5$ & $10$ & $25000$ & $3000$ & $2500$  \\

   &  & & $5$ & $8$ & $24960$ & $2400$ & $3120$  \\
  
  \textbf{Data Pool:} &  & & $5$ & $6$ & $25020$ & $1800$ & $4170$  \\
  
  $n_{tr} \ (n_{pat}) = 173271 \ (15133)$ &  & & $4$ & $4$ & $25000$ & $1200$ & $6250$ \\
  
  $n_{vl} \ (n_{pat}) = 22095 \phantom{8} \ (1892)$&  & & $2$ & $2$ & $25000$ & $600$ & $12500$  \\
   
   \cline{2-8}
   
   $n_{te} \ (n_{pat}) = 20500 \phantom{8} \ (1892)$ & Samples & $32 \times 32$ & $5$ & $6$ & $32400$ & $3000$ & $5400$  \\

   &  & & $5$ & $6$ & $27000$ & $3000$ & $4500$  \\
  
   \textbf{All Labels:} &  & & $5$ & $6$ & $18000$ & $3000$ & $3000$  \\
  
   Epidural, Subarachnoid,  &  & & $5$ & $6$ & $9000$ & $3000$ & $1500$  \\
  
    Subdural, Intraparenchymal, &  & & $5$ & $6$ & $6000$ & $3000$ & $1000$  \\
  
    Intraventricular, No Finding &  & & $5$ & $6$ & $3000$ & $3000$ & $500$ \\
   
    &  & & $5$ & $6$ & $1800$ & $3000$ & $300$  \\
   
   &  & & $5$ & $6$ & $600$ & $3000$ & $100$  \\
   
   \cline{2-8}
   
   & Resolution & $32 \times 32$& $5$ & $20$ & $117168$ & $4000$ & $155-85876$ \\
   
   & (pixels)  & $64 \times 64$& $5$ & $20$ & $117168$ & $4000$ & $155-85876$  \\
   
   &  & $128 \times 128$& $5$ & $20$ & $117168$ & $4000$ & $155-85876$ \\
  
  \hline 
\end{tabular}}
\caption[benchmark setting]{\textbf{All benchmark settings.} Each setting refers to the number of images and radiology findings that are present in the real training, validation and test sets. After GAN training, the synthetic datasets are generated by conditioning on the same label sets, resulting in equivalent data folds. \textit{Dataset information} summarizes the total amount of data in each dataset after preprocessing. $m_{labels}$ refers to the number of labels. $m_{label \ comb}$ refers to the number of unique label combinations. Note that the number of label combinations can be smaller than the number of labels if labels co-occur. For example with three binary labels there might be only two combinations $[1,1,0]$ and $[0,0,1]$, because the first two labels are both either $0$ or $1$. For simplicity and clarity we did not summarize co-occurring radiology findings into a new label combination e.g. turn the aforementioned example into $[1,0]$ and $[0,1]$ with a new binary label "pneumonia \& support devices". $n_{tr}$, $n_{vl}$, $n_{te}$ refers to the total number of training, validation and test samples. $n_{per \ label \ comb}$ refers to the number of training samples per unique label combination.}
\label{tab:1}
\end{table}

\subsection{GAN model development} 

\subsubsection{prog-GAN} 

We used the prog-GAN model as originally proposed in \cite{karras2018progressive}, as it is still regularly used for generating medical images \cite{ali2019data, han2019breaking}. The input of the generator is a concatenation of the $512$ dimensional random normal noise vector $z$ and the label information $y$. Each resolution block is composed of two $3\times3$ convolutional layers followed by Leaky-ReLU activation functions and pixel-wise feature vector normalisation. For networks operating at up to $32 \times 32$ pixels, the generator operates at constant $512$ feature channels. At higher resolution, the number of feature channels is halved with the final convolution layers of the $64\times64$ and $128\times128$ block. The discriminator consists of the same resolution blocks in the opposite order and without pixel-wise feature vector normalisation. When operating above $32 \times 32$ spatial resolution, the first convolutional layer in each block doubles the number of feature channels. In the final layer of the discriminator, the mini-batch standard deviation across all channels is added as an additional feature channel to increase variation. Between resolution blocks, nearest neighbour upsampling doubles the generator's resolution, and downsampling by average pooling halves it inside the discriminator. At each operating resolution, $1 \times 1$ convolutional layers project the number of feature channels to and from the image space, which allows to smoothly interpolate between consecutive levels of detail during progressive growth. All weights in the network are dynamically scaled with a variant of He's initialiser \cite{he2015delving} at each optimisation step to stabilise training. The Wasserstein GAN with gradient penalty loss function is used \cite{gulrajani2017improved}. An additional auxiliary classifier loss term is added to both the generator and discriminator \cite{odena17a} for conditioning. The discriminator is not only trained to classify whether input images are real or fake, but to additionally predict the label. The softmax cross-entropy loss between true and predicted labels for both real and fake images is added to the discriminator loss function, while the same loss but only for fake images is added to the generator loss. We analysed several hyper-parameter settings, mainly different batch sizes, learning rates, number of feature channels and optimiser settings, but we determined that the original parameters proposed in \cite{karras2018progressive} performed best. We began training at a spatial resolution of $8 \times 8$ pixels, which we determined to be the lowest resolution at which meaningful information is still visually apparent in downsampled images. Each transition and stabilisation phase at a resolution of $32 \times 32$ pixels lasted until the discriminator had seen $1.4$M real images, which corresponded to $1.4$M fake images as the number of discriminator updates per generator step is $n_{critic}=1$. At a resolution of $64 \times 64$ and $128 \times 128$ pixels, we reduced the number of real images per phase to $1$M. 

\subsubsection{cpD-GAN}
\label{sec:met_cpd}
\begin{figure}[h]
\center
\includegraphics[width=.6\linewidth]{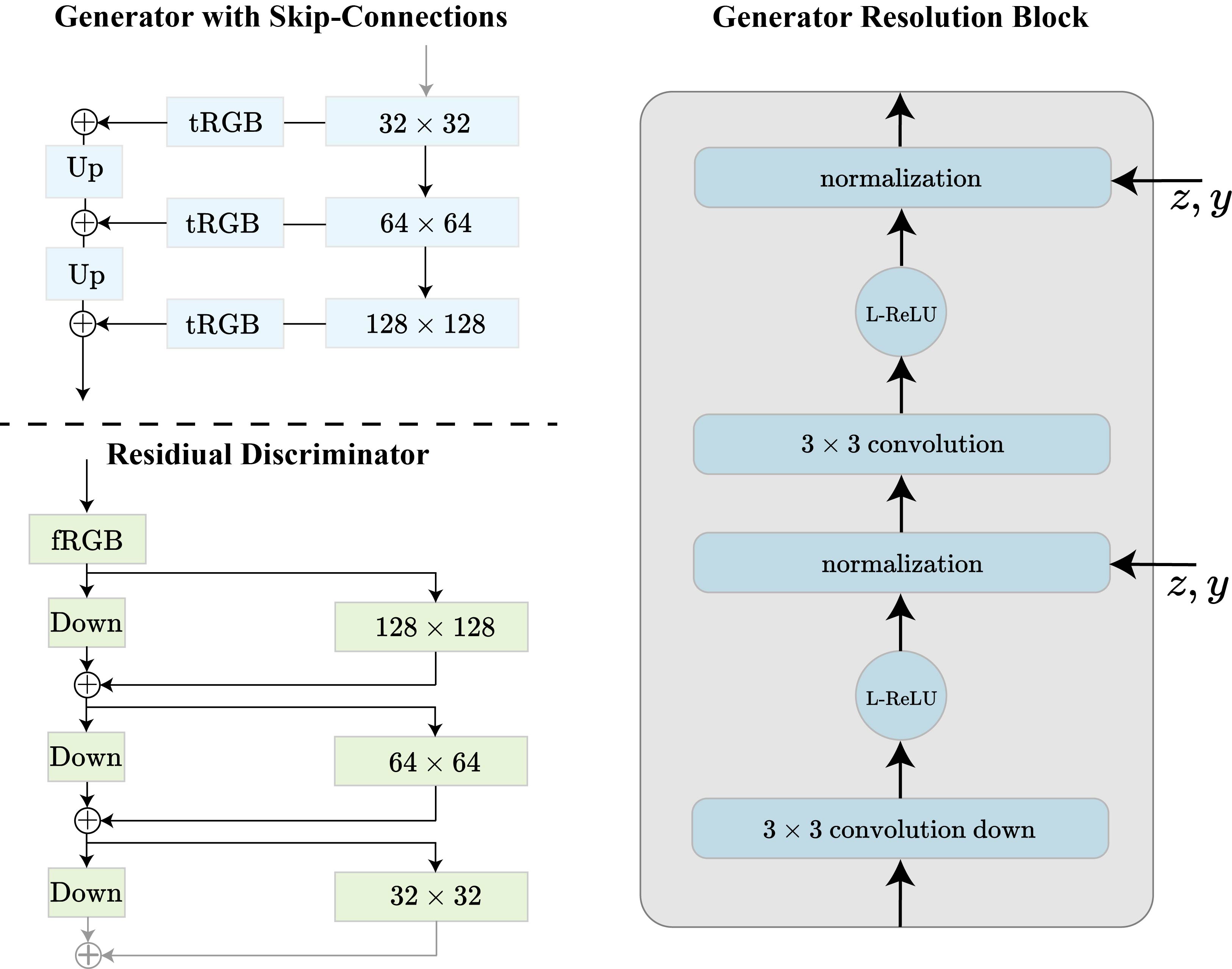}
\caption[Network architecture and generator block]{\textbf{Network architecture and generator block.} \textbf{Left:} In the generator, output skip connections in the image feature space are included after each resolution block, while the discriminator blocks have residual connections. \textit{Up} and \textit{Down} refer to nearest neighbour upsampling and downsampling by average pooling while  \textit{tRGB} and \textit{fRGB} refer to the $1 \times 1$ convolution mappings to and from the image space. \textbf{Right:} The first convolution in each generator block doubles the spatial resolution via nearest neighbour upsampling and reduces the number of feature channels (if needed). Each pixel-wise feature vector normalisation layer is conditioned on the label information $\bold{y}$ and random normal noise vector $\bold{z}$. The Leaky-ReLU non-linearity is used as an activation function.} 
\label{fig:s1}
\end{figure}

We developed the cpD-GAN based on the prog-GAN with several important improvements that we highlight below. Please see above or \cite{karras2018progressive} for details on the architecture and methods if not explicitly stated. Inspired by Style-GAN \cite{ Karras2019AnalyzingAI,karras2019}, we dropped progressive growth as we observed that it was not necessary for stable training. This allowed us to experiment with new architectures, where output skip connections within the image feature space of the generator and standard residual connections in the discriminator improved the performance the most. We achieved significantly lower $\overline{AUC}_{\rm real}-\overline{AUC}_{\rm syn}$ scores when replacing the auxiliary classifier conditioning with a projection based discriminator: In the last discriminator layer, the inner product between the label vector $y$ and the feature vector is computed as the final output, resulting in a conditioning mechanism that respects the role of the conditional information in the underlining probabilistic model \cite{miyato2018cgans}. Inspired by conditional batch normalisation \cite{harm2017}, we modified the pixel-wise feature vector normalisation after each generator convolution by conditioning it on a label and noise dependent scaling and bias parameter: 

\begin{equation}
    b^i_{x,y} = \frac{a^i_{x,y}}{\sqrt{1/N \sum_{j=0}^{N-1}(a^j_{x,y})^2+\epsilon}} \cdot \gamma^i + \beta^i
\end{equation}

where $a^i_{x,y}$ and $b^i_{x,y}$ are the original and normalised feature of channel $i$ in pixel $(x,y)$ and $\epsilon = 10^{-8}$. The scaling parameter is defined as $\gamma = W_1[z;y] + b_1$ and the bias parameter as $\beta = W_2 [z;y] + b_2$, where $W_i$ and $b_i$ are trainable weight matrices and vectors, while $[z;y]$ refers to the vector concatenation of the random normal input noise $z$ and label $y$. Figure \ref{fig:s1} shows the overall model structure and a detailed description of a generator resolution block. 

We evaluated various loss functions, such as the logistic GAN loss with and without R1 or R2 regularisation, the hinge loss with and without gradient penalty, or the non-saturating GAN loss \cite{mescheder18a}, but the Wasserstein loss with gradient penalty worked best. Replacing a specific convolutional layer in both the generator and discriminator by a normal, sparse, or non-local self-attention layer \cite{Wang2018NonlocalNN} did not improve performance. Neither consistency regularisation \cite{zhang2019consistency}, nor matching the gradients of an auxiliary classifier by minimisation of the cosine-distance when predicting the labels of fake and real images resulted in better scores. We analysed many hyper-parameters, among others the number of feature channels, batch sizes, and learning rates. The performance peaked for $512$ feature channels, a batch size of $256$ and learning rates of $0.005$ with one discriminator update per generator update $(n_{critic}=1)$. 

\subsubsection{BIGGAN (discarded)} 
\label{sec:biggan}

The third model that we analysed in detail is largely based on the normal BIGGAN implementation \cite{brock2018large}, with some elements of the self-attention GAN \cite{ZhangGMO19}. However, the implementation did not generalise across different benchmark settings which is why we excluded it from the results and discussion section. In the generator, each block has residual connections and is made up of two $3\times3$ convolutional layers (the first halves the number of feature channels), with ReLU non-linearities followed by conditional batch normalisation and nearest neighbour upsampling layers in between. The $120$-dimensional random normal noise vector $z$ is split, concatenated with the label vector $y$, and fed as input to the initial fully connected generator layer and every residual block. The output layer of the generator consists of batch normalisation, a $3\times3$ convolutional layer and $tanh$ non-linearity. In the conditional projection based discriminator, residual blocks are built in the opposite way, without batch normalisation and with average pooling for downsampling. In both the generator and discriminator a self-attention layer replaces the residual block at the second highest spatial resolution. To stabilise training spectral normalisation, along with orthogonal weight regularisation is applied to all weights \cite{miyato2018spectral}. Prior to the label projection embedding in the discriminator, global sum pooling is performed. We investigated a large amount of different loss functions, feature channel numbers, batch sizes, learning rates and discriminator updates per generator update. Even after performing extensive experiments we could not find a model that generalised across the different dimensions of the benchmark settings, often resulting in training collapse, high FID scores or large $\overline{AUC}_{\rm real}-\overline{AUC}_{\rm syn}$ scores. In our only stable training setting for the resolution at $32 \times 32$ pixels, we used a combination of the hinge loss for the discriminator and Wasserstein loss for the generator, a batch size of $256$ with a maximum of $256$ feature channels, learning rates for the generator and discriminator of $0.01$ and $0.04$ and two discriminator updates per generator update $n_{critic}=2$. 

\subsection{GAN training} 
\label{sec:gan_train_fid}

We used the Adam optimiser for all GAN models and the hyper-parameters as proposed in \cite{karras2018progressive, brock2018large}, except for the learning rates that we fine-tuned as mentioned in \ref{sec:met_cpd}. We stopped training in all settings when the Fréchet Inception Distance (FID) between $10,000$ real and synthetic images converged. The FID score is a commonly used metric to compare the visual quality between images synthesized by generative models and the real training data and tracking it allows for an unbiased GAN training evaluation \cite{heusel2017gans}. A low FID score means that the visual quality of the synthetic images is close, compared to the set of real images. More precisely, for both sets of images the coding layer representations of a pre-trained Inception model (v3) are extracted to obtain vision-relevant features. The sets of features are summarized as a multivariate Gaussian by calculating the mean $(\bold{m})$ and covariance $(\bold{C})$ and the FID score is computed as:

\begin{equation}
    d^2((\bold{m},\bold{C}), (\bold{m}_w,\bold{C}_w)) = ||\bold{m}-\bold{m}_w||_2^2 + \text{Tr}(\bold{C}+\bold{C}_w - 2(\bold{C}\bold{C}_w)^{1/2})
\end{equation}

where $(\bold{m}_w,\bold{C}_w)$ are the mean and covariance of real image feature representations, while $(\bold{m},\bold{C})$ refers to the statistics for synthetic images and $\text{Tr}$ refers to the trace, the sum of the diagonal elements of the matrix.
We ran all models for a minimum number of steps, until the discriminator had seen as many real images as the prog-GAN discriminator after the final stabilisation phase. At a resolution of $32 \times 32$ pixels, each progressive phase lasted until the discriminator had seen $1.4$M real images, resulting in a minimum number of $7$M real images for the other models. For $64 \times 64$ and $128 \times  128$ pixels, we lowered the number of images per phase to $1$M, resulting in a minimum number of images of $7$M and $9$M, respectively. At this point, we computed the FID score after every $400$T real images, and if there was no improvement for two consecutive evaluations, we stopped training. We stopped all repetitions for each experiment at the same step as the first model to get comparable results. For the number of classes benchmark the convergence point for all experiments on both datasets was between $7.0$M and $9.6$M real images. For the number of samples per class benchmark the FID convergence occurred between $7.2$M and $10.8$M real images for sample numbers above $1500$. Below that it took between $14$M and $16$M real images, likely due to the small amount of training data and label overfitting effects. At a resolution of $64 \times 64$ and $128 \times 128$ the GAN models converged between $9$M and $11$M and $12$M and $14$M respectively. At a resolution of $256 \times 256$ pixels for brain scans and x-rays and at $512 \times 512$ pixels for x-rays, the GAN models converged at around $7.5$M and $8.6$M, respectively.

\subsection{Predictive model development and training} 

In all settings, we used a pre-trained densenet-121 convolutional neural network as the predictive model \cite{huang2017densely}. In dense convolutional neural networks for each layer in every dense block, a concatenation of the feature-maps of all preceding layers are used as inputs, and its own feature-maps are used as inputs into all subsequent layers. Each block in the network consists of a number of bottleneck and composition layers, which both have batch normalisation and ReLU non-linearities. In each bottleneck layer the number of feature channels is reduced with a $1\times 1$ convolution, before performing a $3 \times 3$ convolution in each composition layer. Between the dense blocks, the model consists of transition layers with $1 \times 1$ convolutions and average pooling operations. The dense structure allows for significantly deeper architectures without experiencing vanishing gradients, or an exploding number of weights. After the final global average pooling layer, we added a randomly initialised fully connected layer with sigmoid activation for classification with the binary cross-entropy loss. The densnet-121 architecture amounts to $117$ convolution, $3$ transition and $1$ classification layer. We resized the input images to match the densenet-121 spatial input resolution of $224 \times 224$ pixels. To make training as similar as possible across different benchmark settings, we used a maximum number of $5000$ images per epoch with a batch size of $48$. In settings where the total number of samples is below $5000$, the number of images per epoch is accordingly lower. After each epoch, we computed the area under the receiver operating characteristic curve (AUROC) for each label in all validation data samples. We reduce the initial learning rate of $0.0001$ by a factor of $10$ if the mean validation AUROC ($\overline{AUC}_{val}$) across all labels does not improve after two consecutive epochs (patience of $2$). If the $\overline{AUC}_{val}$ does not improve for a patience of $3$ epochs, we stopped training. To compute delta scores, we tested all models on the held-out, real data test set. 

\subsection{Statistical tests for benchmark} 

We repeated every experiment of our benchmark with at least four different random initialisation of the entire training and evaluation pipeline, allowing us to compute the standard deviation for each setting across repetitions. This is necessary as different parameter initialisation resulting from different random seeds can substantially impact the training of deep learning systems. For the number of classes benchmark, we repeated the cpD-GAN training and subsequent synthetic classification as well as the real data classification for $10$ different random initialisation for both datasets at the extrema: For $20$ and $2$ classes for the chest x-rays and $10$ and $2$ classes for brain CT scans. Subsequently we performed the one-sided, parametric-free, Mann–Whitney U test on the $\overline{AUC}_{\rm real}-\overline{AUC}_{\rm syn}$ scores between the extrema to determine whether there is a statistically significant difference. We followed the same approach for the samples per class benchmark with $20$ repetitions at different random initialisation: For $5,950$ and $200$ samples per class for the chest x-rays and $5,400$ and $100$ samples per class for brain CT scans. We once again performed the one-sided, parametric-free, Mann–Whitney U test on the $\overline{AUC}_{\rm real}-\overline{AUC}_{\rm syn}$ scores between the extreme settings to determine the statistical significance.

\subsection{Nearest neighbours} 

To analyse differences between our generated medical images when compared to the training data, we computed the nearest neighbours for a set of randomly sampled synthetics. For both datasets, we used the synthetic images generated by the cpD or prog-GAN model at a resolution of $128 \times 128$ pixels with the lowest $\overline{AUC}_{\rm real}-\overline{AUC}_{\rm syn}$ scores. We used the predictive model trained on real data at the same resolution level to find the final dense layer representation for each synthetic image; a $1,024$ dimensional vector. We compute the same representation for all real training images and determine the pair of synthetic and real images for which the cosine distance between the final densenet representations is minimal. Using a measure of similarity in the predictive model's feature space results in a more reliable determination of nearest neighbours that exploits invariances to shifts and rotations within the image space of the chest radiographs or brain scans. 

\subsection{Feature importance} 
\label{sec:feat_imp}

We computed the causal contribution of image neighbourhoods towards the label prediction with the method of \citet{schwab2019cxplain}. More precisely, we successively zero masked regions of $2\times2$ pixels in the input image and computed the new, increased predictive model loss. If the masking of a particular neighbourhood resulted in a significant loss increase, the region had accordingly higher importance. Importantly, there is no learning involved, as we have access to the ground truth labels for all test images. After $12,544$ repetitions all regions of the $224 \times 224$ pixels input images were successfully masked. To determine the feature importance we subtracted the original model loss and normalised the attribution map. Similar causal contribution maps indicate a similar quality and structure of image neighbourhoods for real and synthetic images, leading to predictive models that attribute the same regions with high feature importance. 

\subsection{Reader study} 
\label{sec:met_rs}

We conducted the reader study by asking trained radiologists to label a set of $100$ images for both data modalities at different resolution levels as real or synthetic with a web-based labeling tool \cite{labelbox}. Each set consisted of $50$ randomly sampled real and synthetic images, generated by the best performing cpD-GAN. Participants were told that each individual image was sampled at random to avoid any bias during evaluation, without knowledge about the total number of reals and synthetics. For the chest x-rays at $128 \times 128$ pixels, $11$ radiologists participated, while $9$ radiologists labeled the brain CT sets $128 \times 128$ pixels. For each higher resolution setting ($512 \times 512$ pixels on chest x-rays and $256 \times 256$ pixels on brain CT scans), $5$ radiologists participated in the reader study. From each labeled set we computed the values for true reals $(TR)$, false reals $(FR)$, true synthetics $(TS)$ and false synthetics $(FS)$, to determine the classification accuracy $\text{acc}=\frac{TR+TS}{TR+TS+FR+FS}$. For the lower resolution settings, we performed the one-sided, non-parametric Wilcoxon signed-rank test to assess whether the distribution of accuracies is equal or less than the mean accuracy of a fully random classifier with $\overline{acc}=0.5$ ($50\%$). Due to a smaller number of participants we do not perform statistical tests for the experiments at $256 \times 256$ and $512 \times 512$ pixels. Instead we only report the accuracy and standard deviation. 

\newpage
\subsection{More figures} 

\begin{figure}[h!]
\center
\includegraphics[width=.8\linewidth]{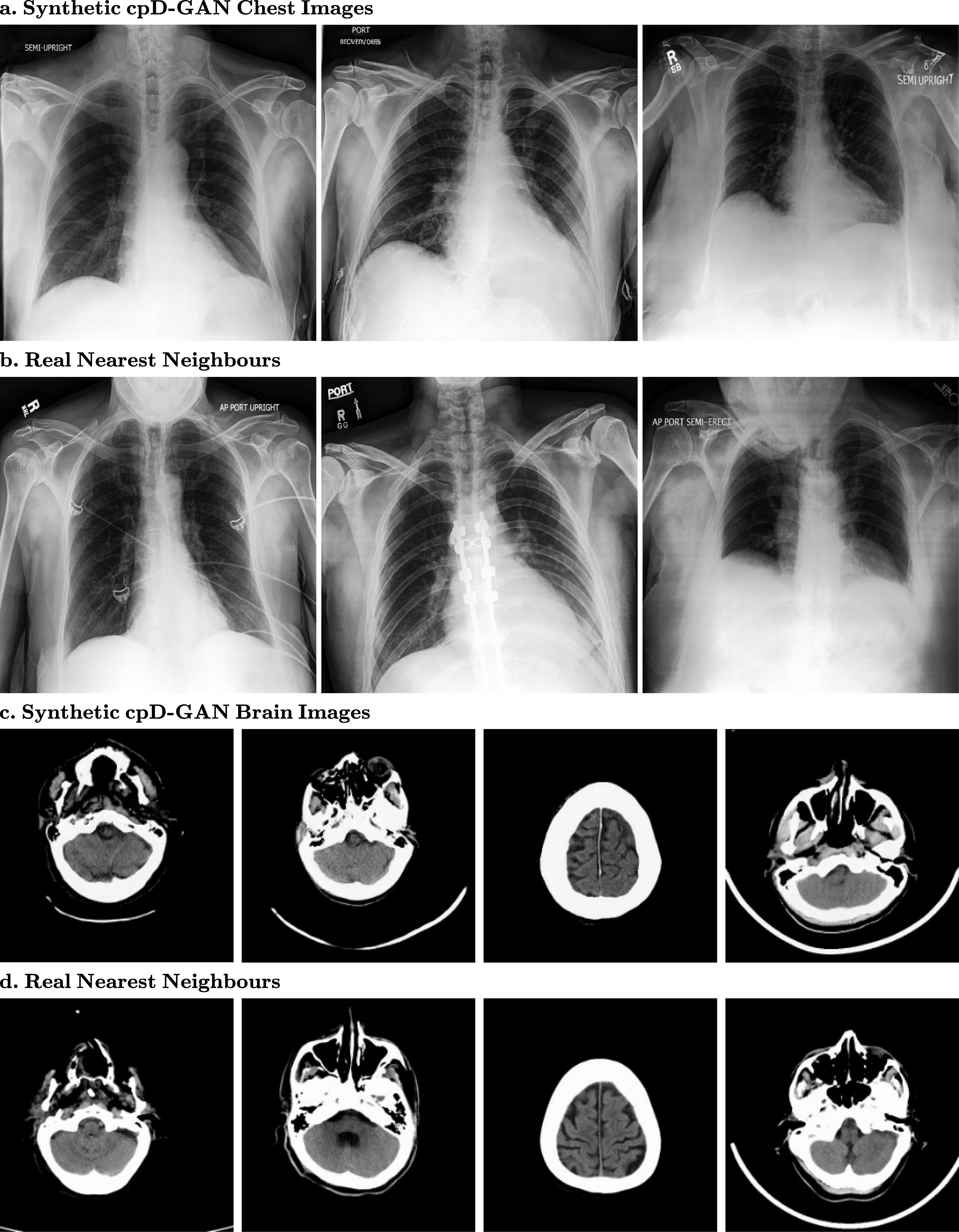}
\caption{\textbf{ More randomly sampled synthetic images generated by the cpD-GAN and real nearest neighbour images from the training.} \textbf{a)} Synthetic chest radiographs at $512 \times 512$ pixels. \textbf{b)} Nearest matching real images found in the chest radiograph training set. \textbf{c)} Synthetic brain computed tomography (CT) scans at $256 \times 256$ pixels. \textbf{d)} Nearest matching real images found in the brain CT training set.}  
\label{fig:s3}
\end{figure}

\clearpage 

\newpage 

\subsubsection{prog-GAN}

\subsection{Details on reader study}
\label{sec:met_conf}

The confusion matrices and accuracies from the reader study at a resolution of $128 \times 128$ pixels are shown in Table \ref{tab:2} and \ref{tab:3}. For the brain CT scans at $256 \times 256$ pixels the results are shown in \ref{tab:4} and for chest x-rays at $512 \times 512$ in \ref{tab:5}.

\begin{table}[h!]
\centering
\begin{tabular}{l|l|c|c|c}
\multicolumn{2}{c}{}&\multicolumn{2}{c}{\textbf{Radiologist}}&\\
\cline{3-4}
\multicolumn{2}{c|}{}&Real&Synthetic\\
\cline{2-4}
\textbf{Actual}& Real & $TR=25.6 \ ( \pm 7.1)$ & $FS= 24.4 \ ( \pm 7.1)$ \\
\cline{2-4}
& Synthetic & $FR= 31.0 \ (\pm 8.2)$ & $TS=19.0 \ (\pm 8.2)$\\
\cline{2-4}
\end{tabular}
\newline
\newline
\newline 
\centering
\textbf{Accuracies of Radiologist Labels}
\begin{tabular}{|l|l|l|l|l|l|l|l|l|l|l|}
\hline
  $0.45$ & $0.52$ & $0.45$ & $0.52$ & $0.50$ & $0.25$ & $0.49$ & $0.39$ & $0.40$  & $0.55$  & $0.39$ \\ \hline
\end{tabular}
\caption[Chest radiographs reader study]{\textbf{Chest radiographs reader study at $128 \times 128$ pixels resolution.} \textbf{Top:} Means and standard deviation from $11$ trained radiologists for real and synthetic images: $TR=\text{True Reals}, \ FR=\text{False Reals}, \ TS=\text{True Synthetics}, \ FS=\text{False Synthetics}$. \textbf{Bottom:} Computed accuracies from radiologist labels.}
\label{tab:2}
\end{table}

\begin{table}[h!]
\centering
\begin{tabular}{l|l|c|c|c}
\multicolumn{2}{c}{}&\multicolumn{2}{c}{\textbf{Radiologist}}&\\
\cline{3-4}
\multicolumn{2}{c|}{}&Real&Synthetic\\
\cline{2-4}
\textbf{Actual}& Real & $TR=25.2 \ (\pm 3.5)$ & $FS=24.8 \ (\pm 3.5)$ \\
\cline{2-4}
& Synthetic & $FR=30.3 \ (\pm 5.8)$ & $TS=19.7 \ (\pm 5.8)$\\
\cline{2-4}
\end{tabular}
\newline
\newline
\newline 
\centering
\textbf{Accuracies of Radiologist Labels}
\begin{tabular}{|l|l|l|l|l|l|l|l|l|}
\hline
  $0.51$ & $0.45$ & $0.44$ & $0.44$ & $0.46$ & $0.50$ & $0.48$ & $0.36$ & $0.40$ \\ \hline
\end{tabular}
\caption[Brain CT scans reader study]{\textbf{Brain CT scans reader study at $128 \times 128$ pixels resolution.} \textbf{Top:} Means and standard deviation from $9$ trained radiologists for real and synthetic images: $TR=\text{True Reals}, \ FR=\text{False Reals}, \ TS=\text{True Synthetics}, \ FS=\text{False Synthetics}$. \textbf{Bottom:} Computed accuracies from radiologist labels.}
\label{tab:3}
\end{table}

\begin{table}[h!]
\centering
\begin{tabular}{l|l|c|c|c}
\multicolumn{2}{c}{}&\multicolumn{2}{c}{\textbf{Radiologist}}&\\
\cline{3-4}
\multicolumn{2}{c|}{}&Real&Synthetic\\
\cline{2-4}
\textbf{Actual}& Real & $TR=37.4 \ ( \pm 5.7)$ & $FS= 12.6 \ ( \pm 5.7)$ \\
\cline{2-4}
& Synthetic & $FR= 16.4 \ (\pm 12.5)$ & $TS=33.6 \ (\pm 12.5)$\\
\cline{2-4}
\end{tabular}
\newline
\newline
\newline 
\centering
\textbf{Accuracies of Radiologist Labels}
\begin{tabular}{|l|l|l|l|l|}
\hline
  $0.76$ & $0.46$ & $0.75$ & $0.54$ & $0.93$  \\ \hline
\end{tabular}
\caption[Chest radiographs reader study]{\textbf{Chest radiographs reader study at $512 \times 512$ resolution.} \textbf{Top:} Means and standard deviation from $5$ trained radiologists for real and synthetic images: $TR=\text{True Reals}, \ FR=\text{False Reals}, \ TS=\text{True Synthetics}, \ FS=\text{False Synthetics}$. \textbf{Bottom:} Computed accuracies from radiologist labels.}
\label{tab:4}
\end{table}

\begin{table}[h!]
\centering
\begin{tabular}{l|l|c|c|c}
\multicolumn{2}{c}{}&\multicolumn{2}{c}{\textbf{Radiologist}}&\\
\cline{3-4}
\multicolumn{2}{c|}{}&Real&Synthetic\\
\cline{2-4}
\textbf{Actual}& Real & $TR=31.6 \ (\pm 7.6)$ & $FS=18.4 \ (\pm 7.6)$ \\
\cline{2-4}
& Synthetic & $FR=28.4 \ (\pm 8.3)$ & $TS=21.6 \ (\pm 8.3)$\\
\cline{2-4}
\end{tabular}
\newline
\newline
\newline 
\centering
\textbf{Accuracies of Radiologist Labels}
\begin{tabular}{|l|l|l|l|l|}
\hline
  $0.54$ & $0.46$ & $0.75$ & $0.54$ & $0.37$ \\ \hline
\end{tabular}
\caption[Brain CT scans reader study]{\textbf{Brain CT scans reader study at $256 \times 256$ pixels resolution.} \textbf{Top:} Means and standard deviation from $5$ trained radiologists for real and synthetic images: $TR=\text{True Reals}, \ FR=\text{False Reals}, \ TS=\text{True Synthetics}, \ FS=\text{False Synthetics}$. \textbf{Bottom:} Computed accuracies from radiologist labels.}
\label{tab:5}
\end{table}
\end{document}